\documentclass[pre,twocolumn,nofootinbib]{revtex4-1}
\usepackage{graphics,color,graphicx,amsmath}
\usepackage{xr}
\usepackage{multibib}
\usepackage{xcolor}

\DeclareMathOperator*{\argmax}{argmax}
\DeclareMathOperator*{\argmin}{argmin}

\begin{document}
\bibliographystyle{unsrt}

\newcommand{\br}[1]{\left\langle #1\right\rangle}
\newcommand{\dn}{{\delta_n}}
\newcommand{\dr}{{\delta_r}}
\newcommand{\df}{{\delta_f}}
\newcommand{\dw}{{d_{\rm w}}}
\newcommand{\db}{{d_n}}
\newcommand{\tr}[1]{\textcolor{red}{#1}}

\title{A scaling theory of armed conflict avalanches}
\author{Edward D. Lee,$^{1,2}$ Bryan C. Daniels,$^3$ Christopher R. Myers,$^{2,4}$ David C. Krakauer,$^1$ Jessica C. Flack$^1$}
\affiliation{$^{1}$Santa Fe Institute, 1399 Hyde Park Rd, Santa Fe, NM 87501\\
$^2$Department of Physics, 142 Sciences Dr, Cornell University, Ithaca NY 14853\\
$^{3}$ASU--SFI Center for Biosocial Complex Systems, Arizona State University, Tempe, AZ 85287\\
$^{4}$Center for Advanced Computing, Cornell University, Ithaca, NY 14853
}

\date{\today}

\begin{abstract}
Armed conflict data display scaling and universal dynamics in both social and physical properties like fatalities and geographic extent. We propose a randomly branching, armed-conflict model that relates multiple properties to one another in a way consistent with data. The model incorporates a fractal lattice on which conflict spreads, uniform dynamics driving conflict growth, and regional virulence that modulates local conflict intensity. The quantitative constraints on scaling and universal dynamics we use to develop our minimal model serve more generally as a set of constraints for other models for armed conflict dynamics. We show how this approach akin to thermodynamics imparts mechanistic intuition and unifies multiple conflict properties, giving insight into causation, prediction, and intervention timing.
\end{abstract}

\maketitle

\section{Introduction}
\begin{quotation}
	{\it The battlefield is a scene of constant chaos.}
	\begin{flushright}--- Napol\'{e}on Bonaparte\end{flushright}
\end{quotation}
The unpredictability of armed conflict is cited in the classic texts on warfare, Sun-Tzu's {\it The Art of War}, Lanchester's {\it Aircraft in Warfare}, and Von Clausewitz's {\it Vom Kriege}. %The outcome of armed conflict is governed by a mix of strategy, geography, tactics and chance. 
Despite seeming chaos in the midst of a single conflict, the ensemble of many conflicts displays multiple mathematical regularities including Richardson's law, the scale-free distribution of fatalities in interstate warfare \cite{richardsonVariationFrequency1948,clausetPowerLawDistributions2009}. How Richardson's law and other scaling patterns relate to one another remains unknown \cite{clausetFrequencySevere2007,gillespieEstimatingNumber2017,johnsonSimpleMathematical2013,picoliUniversalBursty2015}, but a framework unifying these and other conflict aspects could facilitate prediction or reveal hidden and spurious causes of surprising outcomes.

We show, by studying the Armed Conflict Location \& Event Data (ACLED) Project \cite{raleighIntroducingACLED2010}, multiple quantitative regularities that we unify in a simple scaling framework \cite{leeEmergentRegularities2019a}. Such regularities are evocative of scaling laws that emerge in disordered, driven physical systems \cite{sethnaDeformationCrystals2017}, in animal societies with long temporal correlations in conflict dynamics \cite{leeCollectiveMemory2017}, elections \cite{fortunatoScalingUniversality2007}, cities \cite{bettencourtOriginsScaling2013}, amongst other social systems \cite{castellanoStatisticalPhysics2009}. We find that law-like behavior at sufficiently long scales in armed conflict data are captured by a randomly branching, armed conflict (RBAC) model. This model has an underlying fractal geography on which conflict ``contagion'' spreads, uniform dynamics of conflict development on this geography, and scale-free fluctuations in virulence, or intensity, between conflicts.

We extract these regularities from the statistics of {\it conflict avalanches}, consisting of spatiotemporally proximate events that have been joined into clusters. The clustered events consist of individual, localized conflict reports in ACLED, which serves as a database for conflict reports worldwide. Given that most of the data is from Africa, we focus on that region. Each conflict report is labeled by type of interaction, geographic location, date, estimated fatalities, involved actors, and other information (see Appendices of reference \cite{leeEmergentRegularities2019a} for more details). Restricting our analysis to armed battles, we use a systematic definition for relating conflict events: we combine all conflict events within separation time $a=128$\,days and separation length $b=140$\,km to generate conflict avalanches. Thus, conflict avalanches define a set of spatiotemporally extended structures characterized by quantitative properties that, complementary to sociopolitical definitions of ``battles'' or ``wars,'' are constructed only using physical measures of distance.

%In general, changing resolution by varying $a$ and $b$ lead to different theories, e.g., in the limit of minuscule separation scales all events are independent. The scales are chosen such that conflict avalanches span a wide range of scales, and the observed scaling laws are insensitive to rescaling in time $a$ and somewhat more sensitive to rescaling in length $b$ perhaps because of hard constraints in the spatial resolution of conflict events and on the boundaries of Africa. For our purposes here, 

\begin{figure}[tb]\centering
	\includegraphics[width=\linewidth]{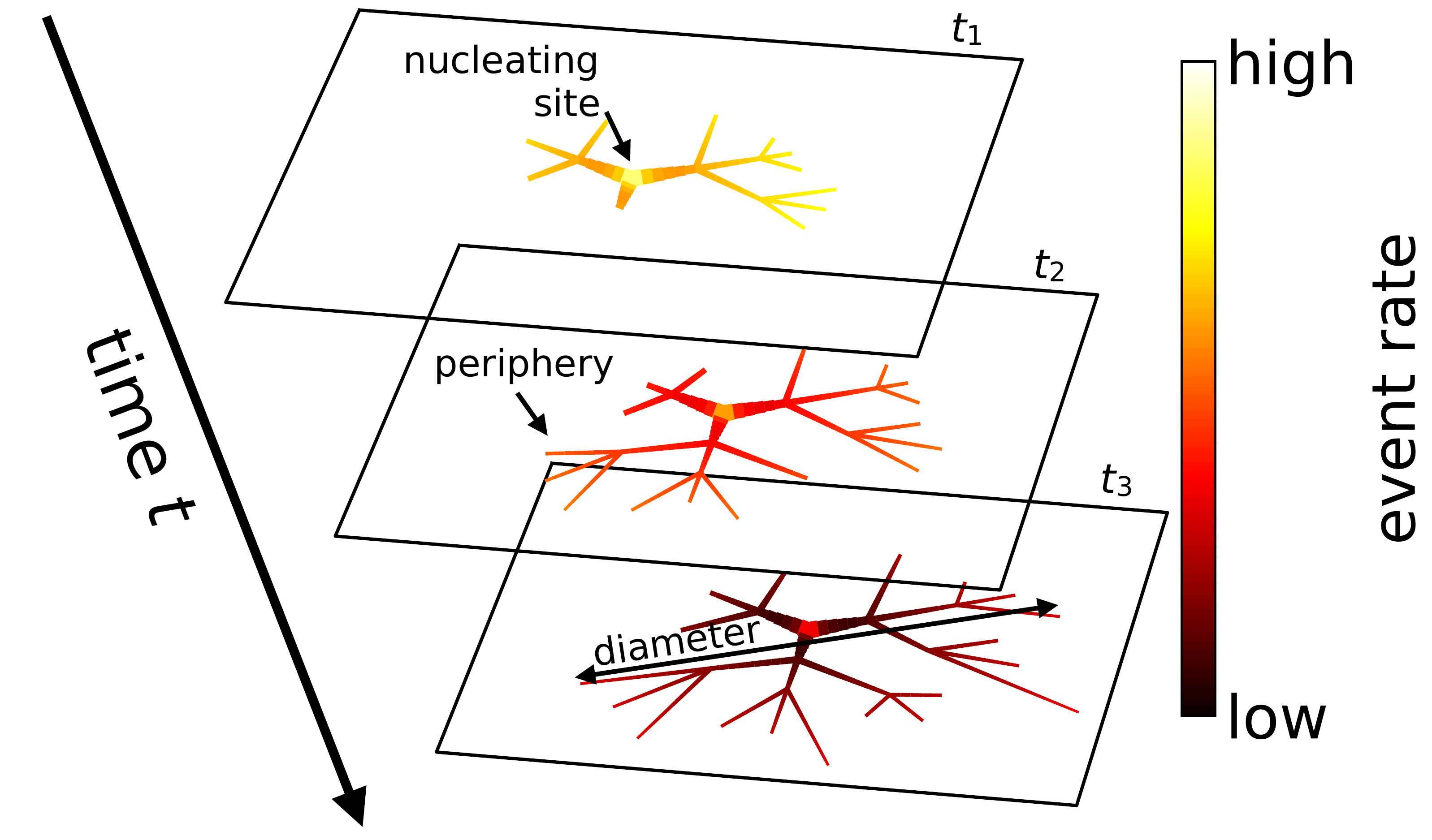}
	\caption{Cartoon of RBAC model. A growing conflict avalanche spreads out in space to new conflict sites in a fractal manner, generating new events at an ever slower rate. As a result, conflict sites near the core tend to have more cumulative events (thick lines) than peripheral sites (thin lines). The rate at which events are generated at the core is higher than that at the periphery, implied by a site growth exponent exceeeding peripheral suppression exponent, $\gamma_r>\theta_r$.}\label{gr:model}
\end{figure}

After having specified $a$ and $b$ (see reference \cite{leeEmergentRegularities2019a} for further details), conflict avalanches can be described by total duration $T$, diameter $L$, infected geographic sites $N$ (a measure of area), fatalities $F$, and number of conflict reports $R$. We discover that conflict properties display power law tails in distribution, scale nonlinearly with duration, and that the exponents characterizing both types of scaling are consistent with one another according to a minimal scaling hypothesis. Over the course of a single avalanche, each of these quantities increases monotonically with time. When they are averaged to generate the dynamical trajectories $l(t)$, $n(t)$, $f(t)$, and $r(t)$, we find that they are invariant under rescaling of the separation time. Taken together, these properties constitute phenomenological scaling variables describing how conflict starts from some epicenter, spreads in space and time, and generates conflict events like fatalities at infected conflict sites such as population centers. With this description as represented in Figure~\ref{gr:model} in mind, conflict avalanches are reminiscent of cascades in other contexts like epidemiology \cite{newmanSpreadEpidemic2002}, neural activity \cite{poncealvarezWholeBrainNeuronal2018,friedmanUniversalCritical2012,fonteneleCriticalityCortical2019}, or stress avalanches in materials \cite{sethnaDeformationCrystals2017}, where universality and scaling provide valid, reduced descriptions of system dynamics. Despite tremendous social, cultural, and ecological complexity, the notion that conflict dynamics likewise conform to a similarly sparse description of conflict contagion is not only an intuitive analogy but one supported by quantitative evidence.

We develop the model in Section~\ref{sec:rbac}, building on previous observations of how conflicts grow to motivate the model \cite{cedermanModelingSize2003,osullivanDominoesDice1996,schutteDiffusionPatterns2011,corralFragilityConflict2020}. We show that the model is consistent with features of the data like functional forms, power law scaling, and exponent relations. For the reader's ease, we provide a table of all the variables discussed in this paper in Appendix Table~\ref{tab:variables} and their estimated values from data, model, and simulation are compiled in Tables~\ref{tab:dynamical exponents} and \ref{tab:exponents}. In Section~\ref{sec:constraints}, we discuss the structure of the model and how it posits a minimal framework for conflict dynamics. Finally, we discuss insights for prediction and intervention in Section~\ref{sec:discussion}.

\begin{figure}[tb]\centering
	\includegraphics[width=\linewidth,clip=true,trim=25 5 25 5]{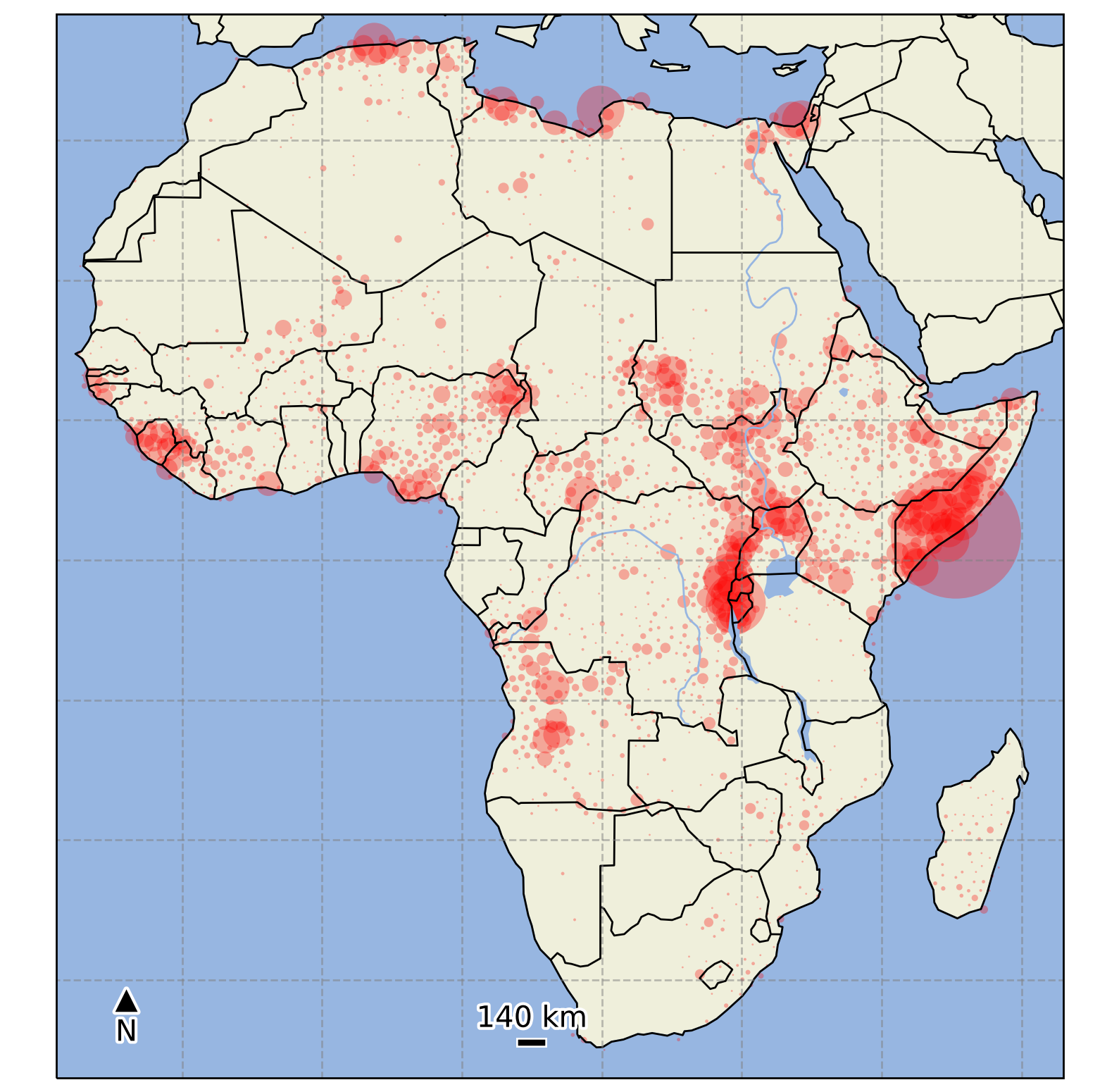}
	\caption{Number of conflict reports averaged over all conflict avalanches per Voronoi region of Africa indicated. Radii of circles are proportional to number of conflict events. Centers of regions are separated on average by separation length $b=140$\,km.}\label{gr:conflict intensity}
\end{figure}

\section{Randomly branching, armed conflict (RBAC) model}\label{sec:rbac}
\subsection{Model dynamics for conflict spread}
We first draw a qualitative outline of our RBAC model. Imagine a big, compact region of length $\Lambda$ that is susceptible to conflict. If conflict breaks out at a central site $x_i$, it ``infects'' neighboring sites through transportation and social networks, growing outwards from the nucleation site $x_{0}$ to cover a set of sites $x\equiv \{x_i\}$, a conflict avalanche of diameter at most $\Lambda$. At each newly infected region (e.g., township, county, province), conflict becomes endemic, generating instability, news reports, and fatalities. Far from the nucleating site, however, conflict potency will be lower as the relevance of distant conflict decays and the density of infrastructure supporting it shrinks (e.g., transportation networks \cite{kalapalaScaleInvariance2006}). Finally, conflict avalanches are characterized by spatiotemporal variation such that some regions or epochs show much more activity, a kind of spatiotemporally embedded {\it virulence} encoded in the intensity of nucleating events. As we see in Figure~\ref{gr:conflict intensity}, that different regions show strongly varying levels of conflict is empirical fact. Deserts, mountains, and oceans show sparse conflict, if any, but such variation might also result from weak government \cite{marshallGlobalReport2008,corralFragilityConflict2020}, technological variation \cite{hussainOpeningClosed2012}, or historical friction between ethnic groups \cite{sealeStruggleSyria1987}.\footnote{We note, in particular, conflict density is not simply proportional to population density though these are quantities are related.} These elements of geography, endemicity, and virulence define the multiple, parallel processes in our model for armed conflict.

\begin{figure}[tb]\centering
	\includegraphics[width=\linewidth]{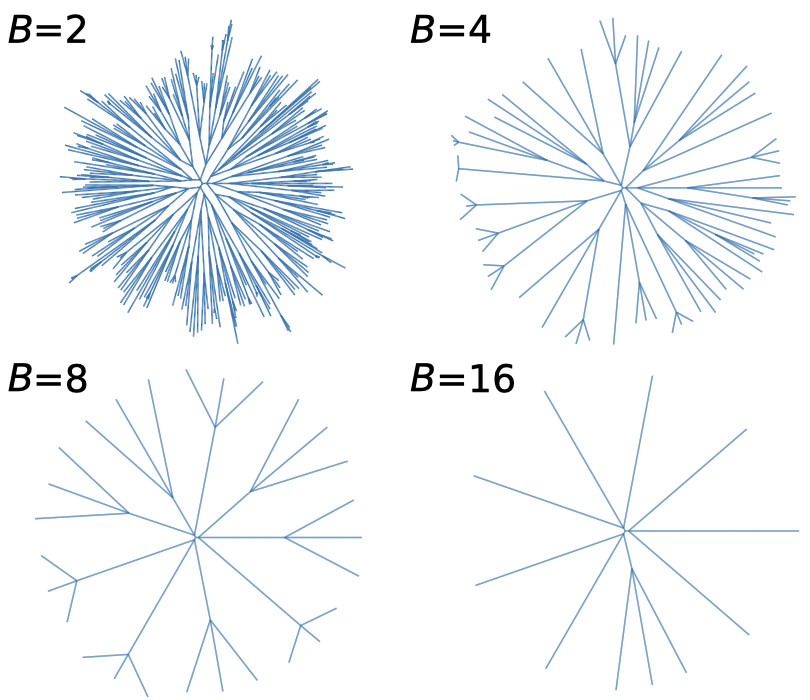}
	\caption{Random variation of ``Nice Trees'' grown for $T=8{,}000$ time steps with branching number $Q=3$ and varying branch length ratios $B$ \cite{burioniFractalsAnomalous1994}. For battles, $B=6.6$. There is one conflict site per unit length.}\label{gr:ntd examples}
\end{figure}

At the core of our model is a randomly branching tree representing the spread of conflict sites at which conflict events occur on the approximately two-dimensional surface of the earth. The tree has branches of average length $B^k$ at generation $k$, each of which give birth to an average of $Q$ branches when they reach their branching points with resulting fractal dimension $\dn = 1+\log(Q)/\log(B)$ as in Figure~\ref{gr:ntd examples}. The increasingly distant branching points of the tree mimic the way road networks become sparse far from highly interconnected cores \cite{kalapalaScaleInvariance2006}. At each time step, a randomly chosen branch is extended by unit length, reflecting the addition of a new conflict site on which conflict reports begin to accumulate. As a result, the time it takes for a site to reproduce---that is, seed another conflict site and further extend the conflict avalanche periphery---increases as the tree becomes larger in a way reminiscent of how battle fronts spread \cite{schutteDiffusionPatterns2011}. These simple dynamics mean that conflict site number grows linearly with time
\begin{align}
	n(t) &= t, \label{eq:n(t)}
\end{align}
having set $n$ to share units with $t$ in our model. Likewise, average avalanche diameter is determined solely by the fractal dimension after sufficient time,
\begin{align}
	l(t) &\sim t^{1/\zeta} = t^{1/\dn}.\label{eq:l(t)}
\end{align}
Eq~\ref{eq:l(t)} also defines the dynamical exponent $\zeta$, which is equivalent to fractal dimension $\dn$ under these minimal single-site growth dynamics. This minimal model capturing geographic spread cannot explain how conflict multiplies at each new infected location as is measured by reports and fatalities. In fact, the measured spatial dimensions for fatalities and reports apparently exceed the dimension in which they live, $d_F>d_R>2$, because of conflict recurrence in fixed areas (Table~\ref{tab:dynamical exponents}). This implies that in order to capture growth in social dimensions of armed conflict, we must account for a separate set of dynamics evolving on top of the geographic lattice.
%\footnote{This fractal structure is a random variation of Nice Trees which are embedded in Euclidean space in way such that random walkers always diffuse normally (not anomalously), examples of which we show in Figure~\ref{gr:ntd examples} for $r=3$ and various values of $b$ \cite{burioniFractalsAnomalous1994}.}

On each site $x_i$ that is infected on day $t_0(x_i)$, conflict becomes endemic and a cascade of conflict events begins. A cascade on site $x_i$ generates conflict reports as well as fatalities, the cumulative numbers of which we track as $r_{x_i}(t)$ and $f_{x_i}(t)$. A phenomenological scaling model for reports at site $x_i$ is
\begin{align}
\begin{aligned}
	&r_{x_i}(t) = \\
	&\left\{\begin{array}{ll}
						v_r(x_i)[t-t_0(x_i)+\epsilon]^{1-\gamma_r}[t_0(x_i)+\epsilon]^{-\theta_r}, & t\geq t_0(x_i); \\
						0, & t<t_0(x_i),
				 \end{array}\right.\label{eq:site scaling}
\end{aligned}
\end{align}
with an analogous equation for $f_{x_i}(t)$. Eq~\ref{eq:site scaling} accounts for site virulence $v_r(x_i)$ modulating local magnitude, growth scaling with exponent $1-\gamma_r$ shared across all conflict sites, peripheral suppression for sites that start later characterized by exponent $\theta_r\geq0$, and a finite rate at all times, $\epsilon=1$. When the growth exponent is at its maximum value $\gamma_r=1$, the new event rate decays quickly, and event count is solely determined by virulence, start time, and peripheral suppression. 

\begin{figure}[tb]\centering
	\includegraphics[width=.9\linewidth]{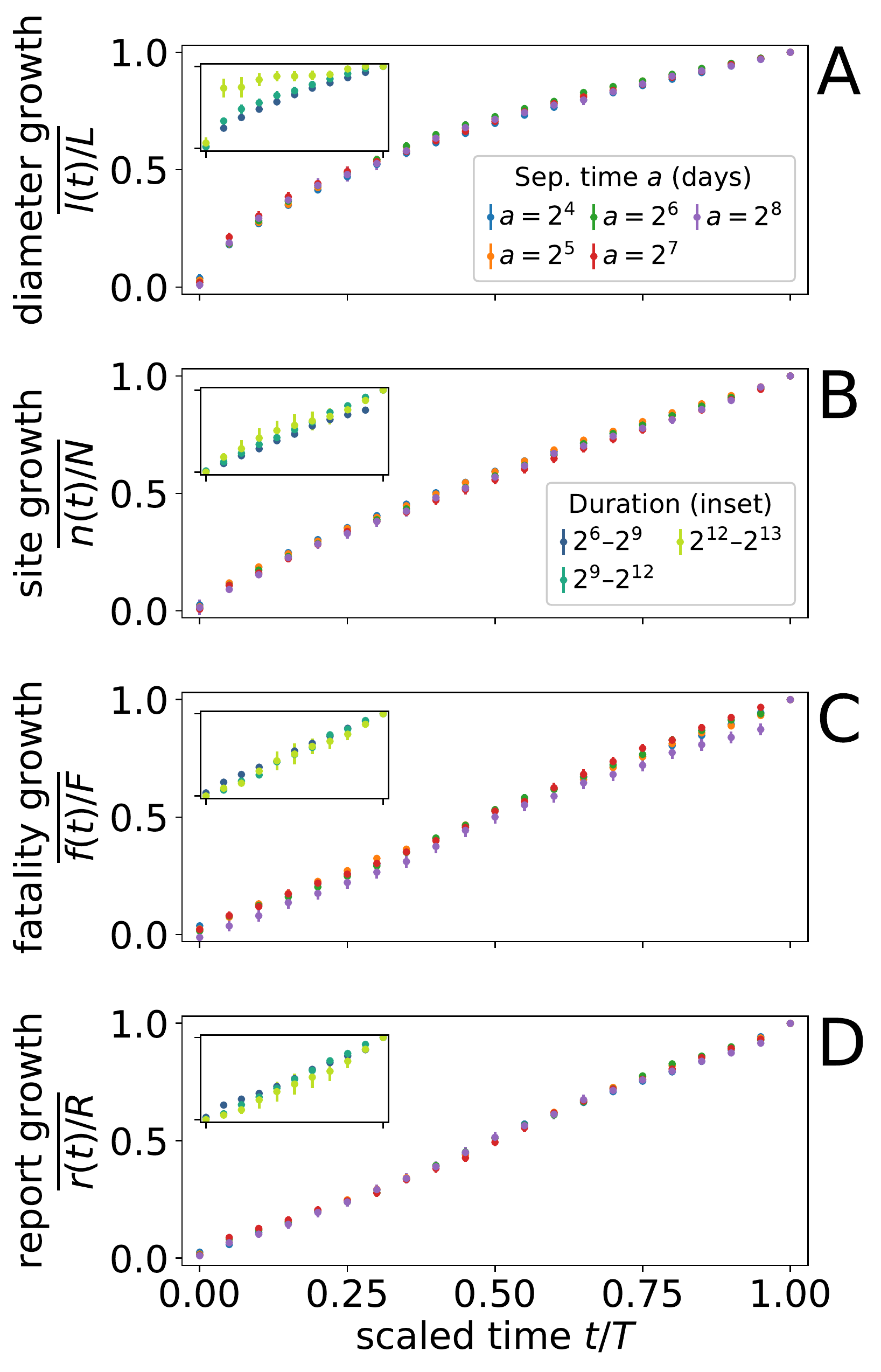}
	\caption{Dynamical scaling collapse for diameter, extent, fatality, and report trajectories under rescaling of separation time $a$. (inset) Scaling collapse for avalanches of different duration. The particularly poor exception is diameter growth $\overline{l(t)/L}$. Data points are few for the longest avalanches, but we find long avalanches saturate the maximum diameter suddenly. Inspecting these avalanches in detail, we find they tend to hit hard boundaries like coastlines and national borders they cannot surpass. In the case of the Tunisian and Libyan revolutions, the aggregation of which is included in the shown average for the longest conflicts, the population is largely confined to the coastline. This suggests for conflict avalanches commensurate with geographic or political boundaries, it is essential to account for such boundaries delimiting their maximum extent.}\label{gr:dynamics}
\end{figure}

By accumulating over the entire extent of the conflict avalanche, we obtain
\begin{align}
	r_x(t) &= \sum_{x_i\in x} r_{x_i}(t).\label{eq:r(t)}\\
\intertext{We expect that at large scales report growth scale with time,}
	r_x(t) &\sim t^{\delta_r/\zeta}\label{eq:r(t) 2},\\
\intertext{a scaling relation that defines the dynamical exponent $\delta_r/\zeta$. In order to proceed with the calculation, we assume that random fluctuations in site virulence $v_r(x_i)$ are uncorrelated with the time at which a site appeared and use a mean-field approximation averaging over conflict avalanche extent, assumptions we verify with data later. Then, Eq~\ref{eq:r(t)} only depends on temporal and spatial scales,}
	r_x(t) &= l(t)^{\dn}V_r(x) \br{[t-t_0(x_i)+\epsilon]^{1-\gamma_r}[t_0(x_i)+\epsilon]^{-\theta_r}},
\intertext{where we have denoted $V_r(x)\equiv \br{v_r(x_i)}$, the expected virulence over a single avalanche $x$, and the typical number of sites $l(t)^{\delta_n}$. With a single site added at every time step, the probability that any randomly chosen conflict site was first infected at time $t_0$ is uniform and}
	 r_x(t) &\sim V_r(x) t^{1-\gamma_r-\theta_r+\dn/\zeta}\label{eq:r(t) 3}
\end{align}
for sufficiently large $t$. Normalizing Eq~\ref{eq:r(t) 3} by $R_x\equiv r_x(t=T)$ to remove dependence on conflict region virulence, we average over $x$ to obtain the universal scaling function
\begin{align}
	\overline{r(t)/R} &= (t/T)^{1-\gamma_r-\theta_r+\dn/\zeta}.\label{eq:norm r(t) collapse}
\end{align}
This presents our first exponent relation for growth in reports using the definition in Eq~\ref{eq:r(t) 2},
\begin{align}
	\delta_r/\zeta &= 1-\gamma_r -\theta_r + \dn/\zeta.\label{eq:r(t) exp relation}
%	&= 2+\gamma_r -\theta_r 
\end{align}
A similar relation holds for fatalities $f$. Taken together, Eqs~\ref{eq:n(t)}--\ref{eq:r(t) exp relation} describe predictions of functional forms and exponent relations that we verify with data.

\begin{table}[tb]
	\caption{Dynamical exponents measured from Battles data, calculated analytically for RBAC model, and estimated from simulation ($K=10^5$ samples). See Figure~\ref{gr:rbac model} for scaling in simulation.}\label{tab:dynamical exponents}\vspace{7pt}
	\includegraphics[width=\linewidth]{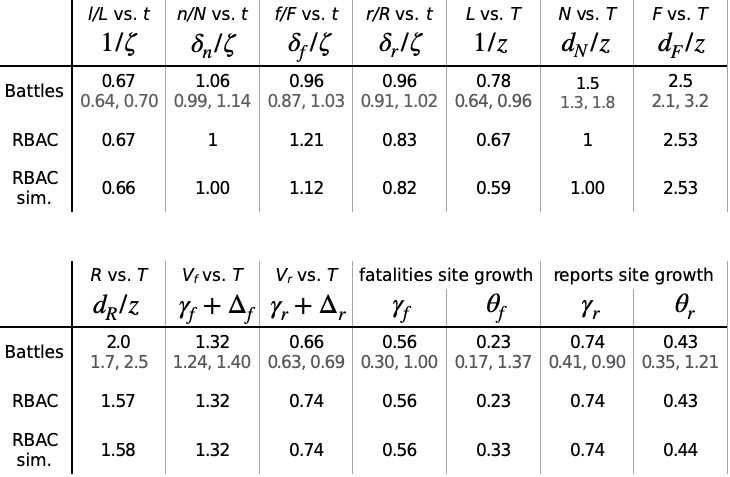}
\end{table}

\subsection{Verifying dynamical model on data}
Using the conflict avalanches that we construct with the data from ACLED as discussed in reference \cite{leeEmergentRegularities2019a}, we check whether or not predictions about universality and self-consistent exponent relations are supported by the data.

As one test of our predicted scaling form for normalized trajectories in Eq~\ref{eq:r(t) 2}, we construct conflict avalanches after rescaling separation time $a \rightarrow 2a$. Under such a change, conflict avalanches will increase in size and duration, although in a way that leaves the normalized functional form unchanged. We show in Figure~\ref{gr:dynamics} over an order of magnitude of rescaling in $a$, the normalized scaling form to be well-preserved, confirming our predictions in Eqs~\ref{eq:n(t)}, \ref{eq:l(t)}, and \ref{eq:norm r(t) collapse} that the dynamical trajectories do not change under temporal rescaling. 

As another test of the dynamical hypothesis, we compare normalized trajectories of short and long conflict avalanches in Figure~\ref{gr:dynamics}. We find that these trajectories largely collapse onto a universal profile---though national and geographic boundaries have an outsize effect on diameter growth for the largest avalanches. Importantly, these ``finite-size'' effects are not prominent after we include avalanche of all sizes, suggesting that scaling of this average is less sensitive to variation in boundary effects across avalanche scales. Taking note of this difference, we take normalized trajectories averaged over avalanches of all durations to measure dynamical scaling exponents $\dn/\zeta$, $\dr/\zeta$, and $\df/\zeta$ shown in Table~\ref{tab:dynamical exponents}. Given these trajectories, we can immediately check if the model exponent $\dn/\zeta=1$ is close to the data $\dn/\zeta=1.06\pm0.05$, which confirms RBAC does indeed imitate the averaged geographic spread of real conflict avalanches across conflict regions and durations.

To check the predicted dynamical exponent expressions for $\dr/\zeta$ and $\df/\zeta$ like in Eq~\ref{eq:r(t) exp relation}, we must measure the site growth exponent $\gamma_r$ and peripheral suppression exponent $\theta_r$. First, we consider how to measure $\gamma_r$. It can be measured directly from observed conflict trajectories for each site as given Eq~\ref{eq:site scaling}. Taking its logarithm, we can fit for some constant $A = \log[v_r(x_i)] -\theta_r \log[t_0(x_i)+\epsilon]$ and for some value of $\gamma_r$ such that
\begin{align}
	\log[r_{x_i}(t)] = A + (1-\gamma_r)\log[t-t_0(x_i)+\epsilon].\label{eq:log rxi}
\end{align}
We leave inside $A$ the unknown combination of random virulence and exponent $\theta_r$ as we discuss in further detail in Appendix Section~\ref{sec:fitting}. 
After constructing conflict sites by taking Voronoi regions inside a conflict avalanche, we estimate $\gamma_r=0.7\pm0.2$ and $\gamma_f=0.6\pm0.3$ (we show the distributions of the exponents in Figure~\ref{gr:gamma estimate}), values we then use to calculate $\theta_r$.

\begin{table}[tb]\centering
	\caption{Exponents for power law distributions measured from Battles data, calculated analytically for RBAC model, and estimated from simulation. For the distribution of sites, the power law tail is statistically distinct from a simple power law \cite{clausetPowerLawDistributions2009}.}\label{tab:exponents}\vspace{7pt}
	\includegraphics[width=\linewidth]{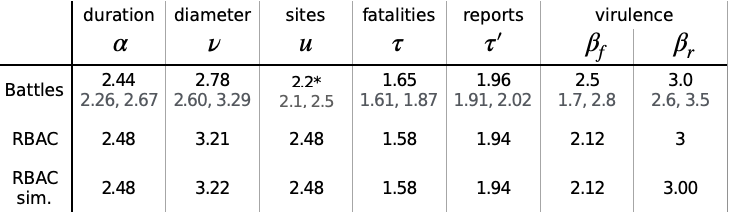}
\end{table}

To measure the decay exponent $\theta_r$, we compute how total activity at a site decays when it starts later in the conflict avalanche by combining decay profiles over all the sites over different avalanches. For a single site, the profiles are
\begin{align}
\begin{aligned}
	r_{x_i}(T)T^{\theta_r+\gamma_r-1} &= v_r(x_i) [1-g(x_i)]^{1-\gamma_r}g(x_i)^{-\theta_r} +\\
	&\qquad\mathcal{O}(\epsilon/T),
\end{aligned}\label{eq:sxi scaling before average}
\end{align}
where we have defined the normalized time at which the site was infected $g(x_i)\equiv t_0(x_i)/T$ and have assumed that the correction to first-order scaling going as $1/T$ is small.\footnote{Caution is warranted at endpoints because the corrections encapsulated in $\mathcal O(\epsilon/T)$ diverge at $g=0$ and $g=1$ as in Eq~\ref{eq:sxi scaling before average}. However, this may not strongly affect the accuracy of measured exponents given that our data set spans only about $\sim8{,}000$ days and almost all our measured avalanches last $T<10^3$\,days.} Taking the average over sites $x_i$ within an avalanche and over conflict avalanches $x$ (denoted by a bar),
\begin{align}
%	\br{r_{x_i}(T)T^{\theta_r+\gamma_r-1}} &\approx V_r(x)[1-g]^{1-\gamma_r}g^{-\theta_r},\\
%\intertext{then considering many different conflict avalanches over regions $x$, }
\begin{aligned}
	\br{\overline{r_{x_i}(T)T^{\theta_r+\gamma_r-1}}} &= \overline{V_r}[1-g]^{1-\gamma_r}g^{-\theta_r} + \\
	&\qquad\mathcal{O}\left(\br{\overline{\epsilon/T}}\right).\label{eq:rxi scaling}
\end{aligned}
\end{align}
Eq~\ref{eq:rxi scaling} describes an averaged conflict event density by the relative time $g$ that has passed, peaking at $g=0$ and sharply suppressed at $g=1$. This particular scaling collapse provides a prediction of how the density of events per site progresses during the course of the avalanche.

\begin{figure}[tb]\centering
	\includegraphics[width=.8\linewidth]{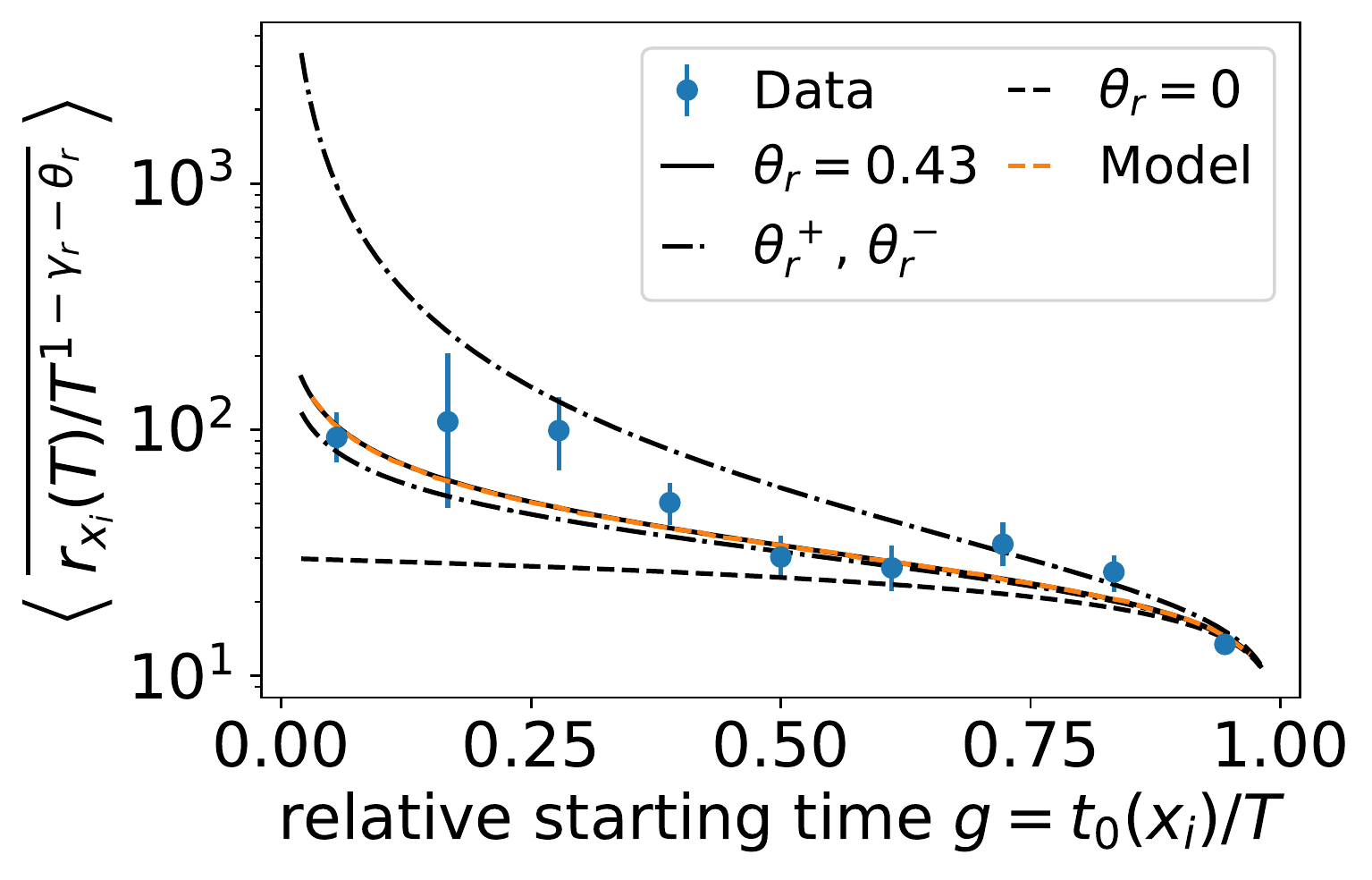}
	\caption{Scaling form predicted in Eq~\ref{eq:rxi scaling} aligns qualitatively with the data given measured $\gamma_r=0.74$. We show bounds on $\theta_r$ corresponding to 90\% bootstrapped confidence intervals as $\theta_r^-$ and $\theta_r^+$ and RBAC simulation (orange). Since for each solution of $\theta_r$ there are corresponding fit parameters from Eq~\ref{eq:theta objective}, the bounding lines for $\theta_r^+$ and $\theta_r^-$ indicate variability in the shape of the curve but not vertical displacement.}\label{gr:theta r}
\end{figure}

Using our estimates for $\gamma_f$ and $\gamma_r$, we use Eq~\ref{eq:rxi scaling} to fit the exponents $\theta_f=0.2\pm0.3$ and $\theta_r=0.4\pm0.3$ with 90\% bootstrapped confidence intervals shown in Table~\ref{tab:dynamical exponents} (see Appendix~\ref{sec:fitting} for measurement details). Importantly, the resulting curves align qualitatively with our predictions as plotted in Figures~\ref{gr:theta r} and \ref{gr:theta f}: the data show an increase in the conflict event rate at sites occurring near the beginning of the avalanche, with strong suppression at the end substantially different from when $\theta_r=0$. With this confirmation, we combine our measured exponents to obtain $1-\gamma_r-\theta_r+\dn/\zeta \approx 0.9$, which is remarkably close to the measured value of $\dr/\zeta=1.06$. Similarly, $1-\gamma_f-\theta_f+\dn/\zeta \approx 1.3$, compared to the best fit estimate from Figure~\ref{gr:dynamics}, $\df/\zeta=0.96$. Though both of these exponent relations are satisfied within bootstrapped confidence intervals, there is substantial  uncertainty in exponent values for conflict site dynamics $\gamma_r$, $\gamma_f$, $\theta_r$, and $\theta_f$ such that the predicted relations are loosely bounded between $\dr/\zeta\in[0,1.4]$ and $\df/\zeta\in[0,1.8]$. That the best fit exponents conform closely to our predicted relations, indeed much closer than the uncertainty suggested by confidence intervals, demonstrates that our formulation aligns well with the dominant features of armed conflict growth. Thus, we find our mean-field formulation of conflict site growth in the RBAC model accurately captures site evolution, peripheral suppression, and tightly satisfies self-consistent exponent relations.

\subsection{Conflict virulence and extinction}
By definition, a conflict avalanche ends when the rate at which new reports $\partial_t r_{x_i}(t)$ are generated falls below some threshold as is set by our separation time $a$. Then, conflict extinction is determined by when the most prolific site at time $t$ falls below rate threshold $C$,
\begin{align}
\begin{aligned}
	C &= \partial_t r_{x_{i^*}}(t) \\
	i^* &= \argmax_i\ \partial_t r_{x_i}(t).
\end{aligned}\label{eq:C}
\end{align}
Given $t$ and looking over sites with starting times $t_0(x_i)$, the rate is dominated by the two peaks at the endpoints with starting times $t_0(x_0)$ and $t_0(x_T)$. As a result, the fastest rate is determined by the relative magnitudes of the exponents $\theta_r$ and $\gamma_r$. Since $\gamma_r>\theta_r$, the rate at the core dominates, and the threshold is met when
\begin{align}
	C \sim V_r T^{-\gamma_r}.\label{eq:V_r for constant threshold}
\end{align}
A universal constant threshold $C$ would imply that $V_r \sim T^{\gamma_r}$. More generally, we might expect that larger conflicts are more difficult to observe because of the ``fog of war'' or if resources for observation are limited such that smaller events do not register as easily \cite{oloughlinPeeringFog2010}. Though our rate threshold is fixed by the separation time, a fluctuating observation threshold could be effectively represented by rate threshold $C$ fluctuating with duration such as
\begin{align}
	C &\sim T^{\Delta_r}. \label{eq:V_r with T}
\end{align}
When $\Delta_r>0$, the threshold increases with conflict duration and thus size, implying that observers are unable to resolve the smaller events unfolding in the conflict.\footnote{On the other hand, $\Delta_r<0$ presents the unlikely possibility that observations become more detailed with increasingly larger conflicts. Such an unrealistic outcome would suggest that this intuitive explanation is flawed, but we find reassuringly the sensible bound $\Delta_r\geq0$ to be satisfied.} In this more general case, the rate threshold condition in Eq~\ref{eq:C} implies
\begin{align}
	V_r \sim T^{\gamma_r + \Delta_r},\label{eq:V_r vs T}
\end{align}
where the exponent $\gamma_r$ describes the decay of conflict event rate at any particular conflict site and exponent $\Delta_r$ describes how the ability to resolve individual conflict events fluctuates with virulence. Similarly, we can construct an argument for fatalities, which likewise leads to a dynamical scaling prediction for conflict virulence of fatalities.

\begin{figure}\centering
	\includegraphics[width=.8\linewidth]{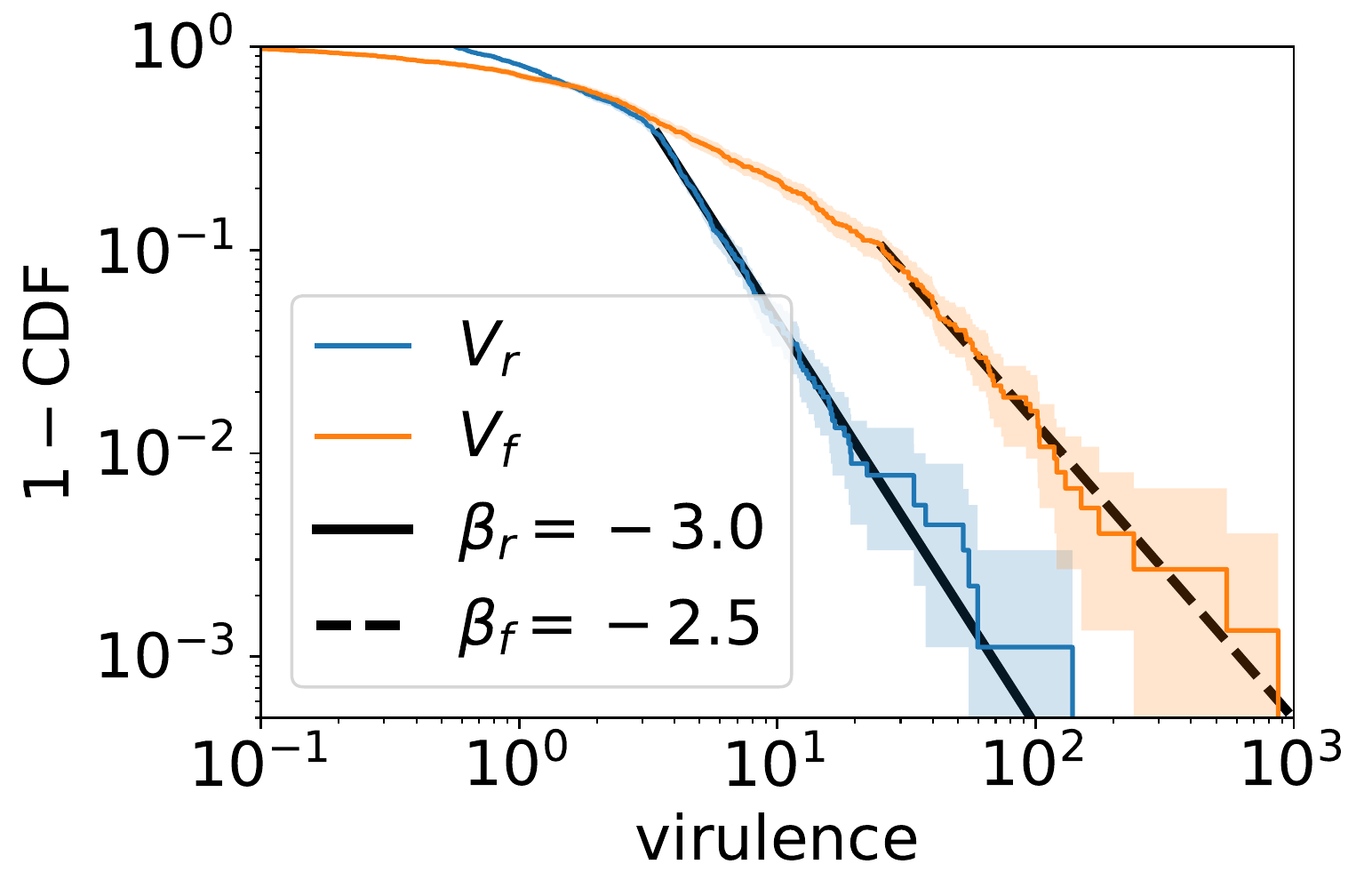}
	\caption{Distributions of average virulence per conflict avalanche $V_r$ and $V_f$ display power law tails whose measured exponents satisfy self-consistent equations derived from the scaling hypothesis ($p>0.8$ compared to standard significance threshold $p=0.1$) \cite{clausetPowerLawDistributions2009}.}\label{gr:virulence}
\end{figure}

This scaling relationship between virulence and duration links local dynamics of conflict growth with conflict avalanche termination, a global property. To take this further, we ask what happens if the distribution of conflict virulence were distributed in a scale-free way, 
\begin{align}
	P(V_r) &\sim V_r^{-\beta_r}.\label{eq:p(vr)}
\end{align}
Fluctuations in $V_r$ would thus induce scaling in conflict duration determined by predicted exponent relation,
\begin{align}
\begin{aligned}
	P(T) &\sim T^{-\alpha},\\
	\alpha &= 1 + (\gamma_r + \Delta_r)(\beta_r-1).
\end{aligned}\label{eq:V_r exponents}
\end{align}
In order to verify this hypothesis, we calculate the virulence for every site in conflict avalanches using our estimates for $\gamma_r$ and $\theta_r$. We show the resulting distributions in Figure~\ref{gr:virulence} for $V_r$ and $V_f$, which both are statistically consistent with having power law tails. From the distributions, we determine $\beta_r=3.0\pm0.3$ and $\beta_f=2.5\pm0.4$. As has been previously noted \cite{leeEmergentRegularities2019a}, the distribution of duration $P(T)$ also displays a power law tail with $\alpha=2.44\pm0.13$. Then comparing virulence with duration $T$, we estimate the dynamical scaling exponent $\gamma_r+\Delta_r = 0.66\pm0.02$. Interestingly, this measurement means that $\Delta_r=0$ is consistent with the data, and that the report rate threshold does not necessarily depend on the intensity of observed conflict. Taking this seriously, we remove an additional parameter by setting $\Delta_r=0$. This is in contrast to the same calculation for fatalities, $\Delta_f+\gamma_f = 1.32\pm0.05$, which implies $\Delta_f>0.3$ given the bound $\gamma_f\leq1$ (see Table~\ref{tab:dynamical exponents}). Such a result suggests that conflict resolution for fatalities fluctuates, a conclusion that aligns with the difficulty of estimating fatalities accurately \cite{raleighIntroducingACLED2010,oloughlinPeeringFog2010}. Reassuringly, these exponents satisfy the predicted scaling relation in Eq~\ref{eq:V_r exponents}, and conflict avalanche extinction aligns with a universal threshold in a way consistent with our a universal separation time scale. Thus, we show the way that we relate virulence and duration, derived from assumptions about scaling and our definition of conflict termination, lead to self-consistent relations satisfied by the data.

\subsection{Scaling framework}
Beyond the scaling of virulence with final conflict duration, the way that the remaining scaling variables---diameter $L$, extent $N$, fatalities $F$, and reports $R$---grow with duration also imply additional power law distributions,
\begin{align}
\begin{aligned}
	P(T) &\sim T^{-\alpha}, & P(L)&\sim L^{-\nu}, & P(N)&\sim N^{-u},\\
	P(F) &\sim F^{-\tau}, & P(R)&\sim R^{-\tau'}.
\end{aligned}\label{eq:distributions}
\end{align}
These are not assumptions but are mathematical consequences of unifying the conclusions in previous sections, and these power laws hold in the data as described at further length in reference \cite{leeEmergentRegularities2019a}. 
The new exponents in Eq~\ref{eq:distributions} are determined by relating site dynamics with total magnitude of conflict avalanche properties after accounting for virulence. Using fatalities as an example, we define the exponent combination $d_F/z$,
\begin{align}
	F \sim T^{d_F/z} \sim V_f T^{\delta_f/\zeta} \sim T^{\gamma_f + \Delta_f + \delta_f/\zeta}.
\end{align} 
Thus, a positive exponent combination $\gamma_f+\Delta_f$ means avalanches grow larger than uniform site dynamics on a branching tree would imply, the excess scaling captured in our model by conflict-site correlations induced by virulence. We calculate from the entries of Table~\ref{tab:dynamical exponents} the contribution of such virulence. Fatalities show strong effects of virulence revealed by the difference $1.0\leq d_F/z - \df/\zeta\leq 2.3$, consistent with $\gamma_f + \Delta_f\approx1.3$, and implying $\Delta_f>0$. Correspondingly with reports, we find that the exponent $\gamma_r=0.74$ accounts for the difference $0.6\leq d_R/z - \dr/\zeta\leq1.5$ such that $\Delta_r=0$, consistent with a fixed conflict termination threshold as noted earlier. Virulence, however, seems to play little to no role in the geographic spread of conflict, $0.2\leq\gamma_n+\Delta_n\leq 0.8$ and $-0.1\leq\gamma_l + \Delta_l\leq0.4$. This observation aligns with our model assumption that virulence is primarily a feature of the social dimensions of conflict but not of geographic spread. 

By connecting the dynamics of conflict growth with the distributions of conflict scaling variables, we unify within a single mathematical model all of these properties and confirm our hypothesis that social growth results from a combination of geographic spread and conflict virulence.

\begin{figure}\centering
	\includegraphics[width=\linewidth]{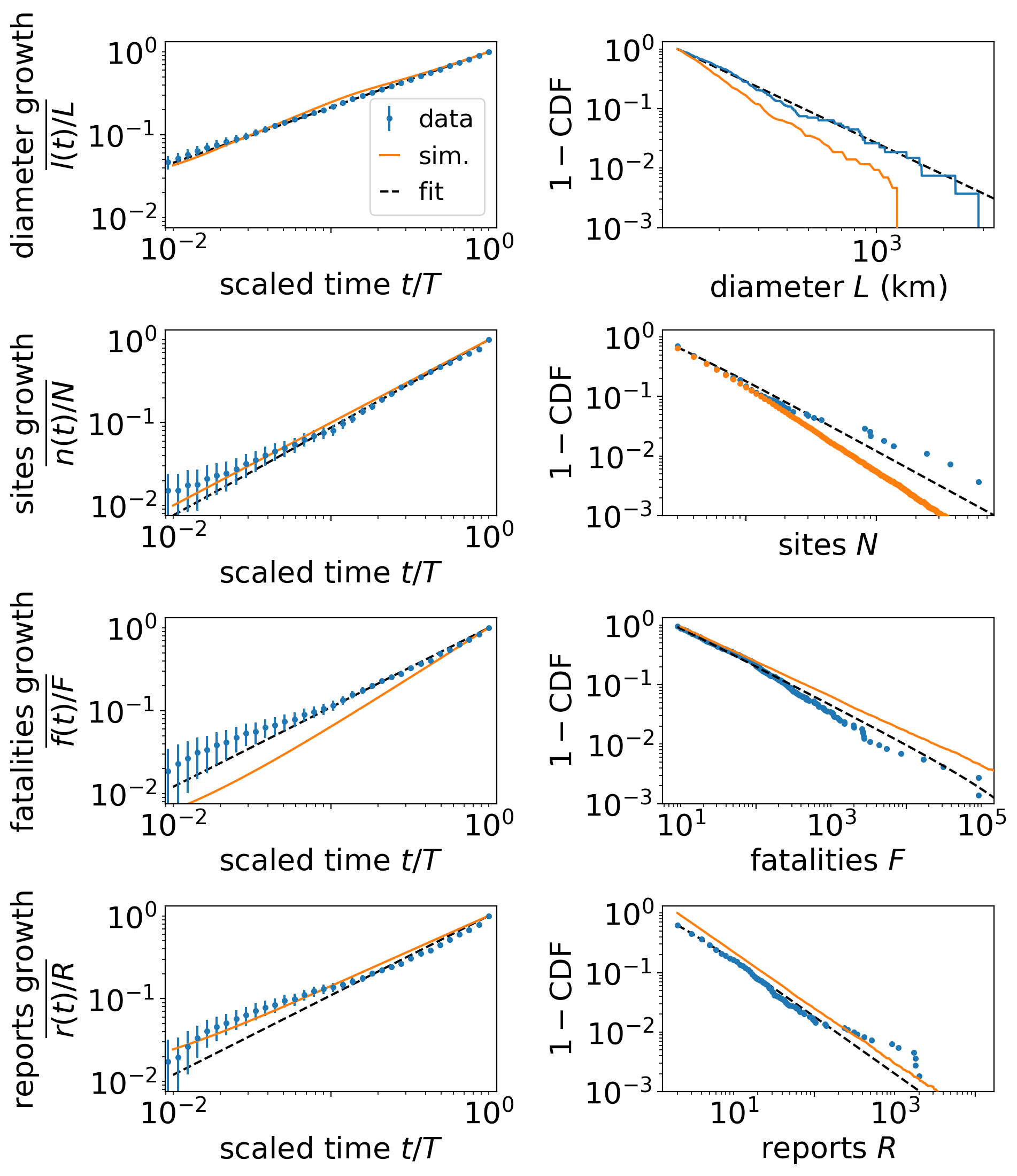}
	\caption{Dynamical scaling and distributions of conflict avalanche scaling variables generated from RBAC compared with data. (left column) Model simulations (orange) closely mimic calculated exponent relations in Table~\ref{tab:dynamical exponents} (dashed black lines) and are similar to scaling in data (blue). Measured dynamical scaling functions are shown after having removed the nonzero intercept at $t=0$ averaged over conflict avalanches with duration $T\geq4$\,days. For $n(t)$, we also require $N>1$ and for $f(t)$ that $F>2$ fatalities. (right column) Distributions of scaling variables with exponents listed in Table~\ref{tab:exponents} align closely. Distributions for both data and RBAC are shown above lower cutoffs and their scales matched such that the lower cutoffs coincide.}\label{gr:rbac model}
\end{figure}

\subsection{Simulation}
As a final check, we simulate the RBAC model. We find close agreement with scaling patterns in the data as shown in Figure~\ref{gr:rbac model} and Tables~\ref{tab:dynamical exponents} and \ref{tab:exponents} (see Appendix Section~\ref{sec:simulation} for further details about the simulation). 

\begin{figure*}[tb]\centering
	\includegraphics[width=.88\linewidth]{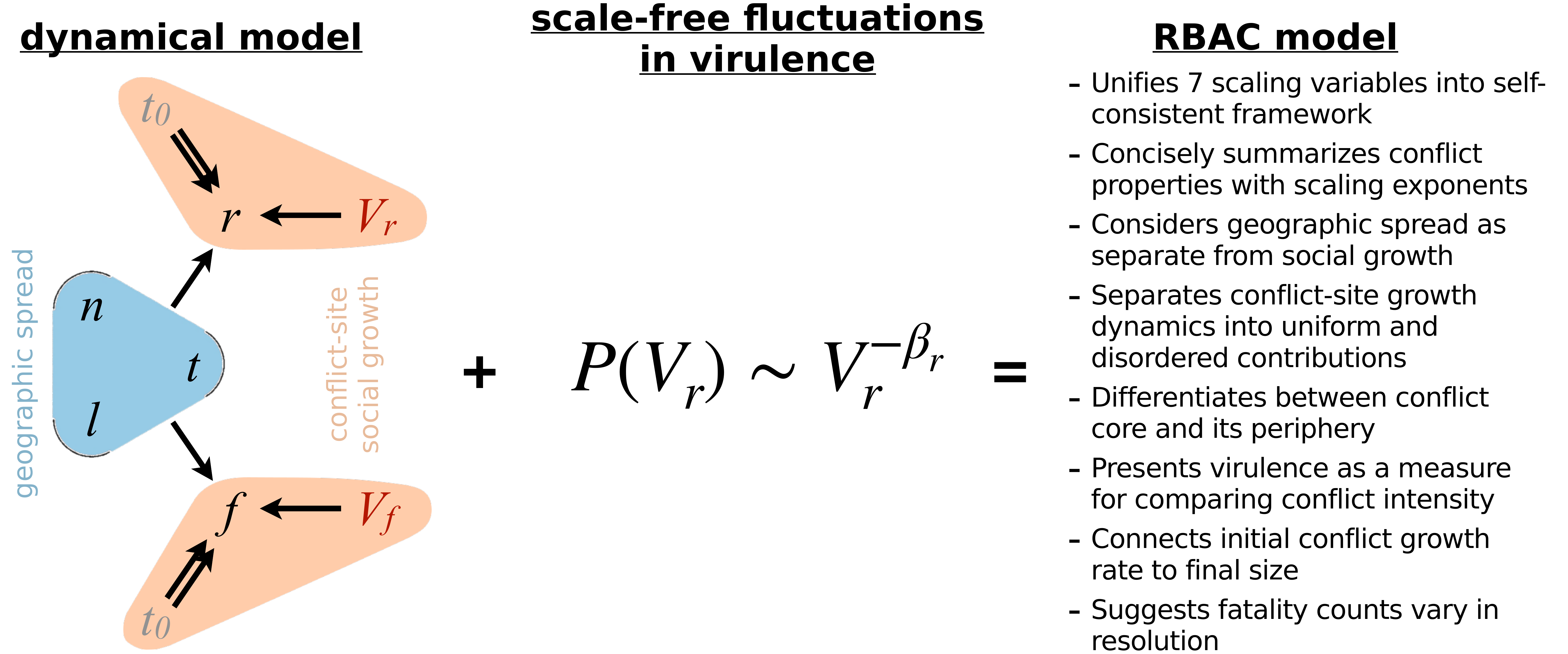}
	\caption{Overview of RBAC model combining a dynamical scaling model with a scale-free distribution of conflict report virulence to generate conflict simulations. (left) Geographic spread of conflict sites involves duration $t$, diameter $l$, and extent $n$, all related by a single exponent. At each conflict site, reports grow in a uniform way, depending only on growth exponent $\gamma_r$, peripheral suppression exponent $\theta_r$, and report virulence $V_r$. To get total report growth $r$, we sum over the geographic extent of the conflict avalanche. Thus, each aforementioned component contributes an additional exponent to $r$ as indicated by the incoming arrows. In contrast to the other scaling variables (black letters), virulence $V_r$ (red) is quenched, or fixed during the conflict avalanche. The variable $t_0$ (gray) indicates when a site first became infected during the course of a conflict avalanche. A similar descriptions holds for fatality growth $f(t)$. (middle) To obtain the scale-free distributions of conflict scaling variables, we further assume a power law form for report virulence distribution $P(V_r)$. (right) All together, the resulting RBAC model leads to several notable features and directions for improving prediction such as detailed in the final two points.}\label{gr:model diagram}
\end{figure*}

\section{A minimal model?}\label{sec:constraints}
Our approach relies on scaling, self-consistency, and simple dynamical hypotheses to build a minimal model that unifies both social and geographic characteristics of armed conflict. Yet, there are sufficiently many components that one might ask if the model is overparameterized. We argue in this section that our model represents a dramatic simplification of the full space of possibilities encompassing 7 scaling variables (i.e., duration, diameter, extent, reports, fatalities, and two types of virulence) and their trajectories. In principle, each of the scaling variables constitutes an independent degree of freedom with infinitely more degrees of freedom for the shape of growth trajectories and their distributions. To specify the functional form of the joint probability distribution relating every such degree of freedom to one another without an informative prior is difficult given the sparse and noisy data available. Instead, we posit a form for the decomposition of the joint probability that is tractable and empirically verifiable starting with assumptions about scaling. %More generally, we show the resulting constraints characterize a wider set of models spanned by uncertainties in determining a few set of features embodied in scaling exponents.

As an example, consider the growth of armed conflict in duration $t$, diameter $l$, and extent $n$. In the most general possible scenario, we have arbitrarily complicated functions relating each pair of variables. However, under our scaling hypothesis, we restrict ourselves to only considering power law forms that correspond to three separate exponents, or degrees of freedom. Under self-consistency and the absence of any additional scaling, the third exponent must be determined in terms of the other two, leading to the relationship $n \sim t^{\dn/\zeta}$ as follows from in Eqs~\ref{eq:n(t)} and \ref{eq:l(t)}. Adding onto this, we assume single-site growth dynamics, which imposes equality of fractal dimension and dynamical exponent, $\dn=\zeta$. Hence, with the case of geographic growth, the combination of scaling, self-consistent exponents, and minimal dynamics compresses an arbitrary number of degrees of freedom into a single degree of freedom captured by the scaling exponent $\dn/\zeta$ that we measure from data (blue triangle in Figure~\ref{gr:model diagram}).

Bringing reports and fatalities into the fold as we show in the leftmost panel of Figure~\ref{gr:model diagram}, our model can be represented as a graph of dynamical scaling variables. In particular, averaged reports growth $r(t)$ is a function of geographic spread, given by $\dn/\zeta$, uniform site dynamics specified by $\theta_r$ and $\gamma_r$, and mean virulence $V_r(x)$. Thus, each aforementioned component contributes an additional exponent to $r$ as indicated by the four incoming arrows. By traversing this sparse graph and taking the exponent relation corresponding to each edge, it is possible to relate every dynamical scaling variable with any other, but note the absence of redundant edges: we have avoided specifying any more edges than necessary to connect all the scaling variables. This dynamical description of conflict growth reduces the open-ended problem of fitting conflict data to specification of a few exponents---to be precise one for the set $t$, $l$, $n$ and two for reports $r_x(t)$ and two for $f_x(t)$---whose relationships align quantitatively with the data. 

The mean virulence $V_r$, however, is unusual as is indicated by its red text color in Figure~\ref{gr:model diagram}. Unlike the other scaling variables in black, it is quenched and so does not change as conflict progresses. Instead, the criterion for conflict extinction relates it to the total duration, linking dynamics with fluctuations in conflict avalanche size. Thus, virulence plays a special role in our theory, driving the intensity of conflict site growth in a uniform way within the context of a single conflict avalanche but displaying scale-free fluctuations across many separate conflict avalanches. This aspect is represented in the midddle panel of Figure~\ref{gr:model diagram} as the power law distribution of virulence $P(V_r)$. With this assumption, we can calculate distributions of all remaining variables using the dynamical scaling relations and obtaining vast simplification. For example, we can construct the distribution of fatality virulence $P(V_f)$ by using the dynamical scaling relations $V_f\sim T^{\gamma_f+\Delta_f}$ and $V_r \sim T^{\gamma_r}$, which imply $V_f\sim V_r^{(\gamma_f+\Delta_f)/\gamma_r}$ and thus a power law form for the distribution $P(V_f)$. Taken together, these components---uniform growth dynamics, scale-free fluctuations in virulence, and avalanche extinction below some threshold rate---compose a set of mathematical relationships between measurable conflict properties that sparsely relate the many aspects of conflict. Beyond our model, these scaling relations serve as constraints delimiting the set of conflict models that, if specifying many further microscopic details and proposed mechanisms for conflict propagation, must still hew to the regularities that we find in the data.

\section{Discussion}\label{sec:discussion}
That the complex tangle of armed conflict reveals strong regularities at large scales is truly remarkable. As one notable example that might have led us to anticipate the opposite, consider the conflict avalanche spanning Tunisia and Libya \cite{leeEmergentRegularities2019a}. This outbreak of civil wars, which was part of the Arab Spring, clearly adheres to the geometry of the coastline given the density of population there. In contrast with other conflicts, this war began with the end of dictatorship and devolved into infighting amongst multiple militias seeking control over land, natural resources, and government \cite{lynchArabUprising2013}. Furthermore, it is difficult to refute the argument that geography plays a defining role in this conflict avalanche's spread. Yet, in the face of many such particulars, the statistics that emerge from the ensemble display highly regular, emergent properties such self-consistent power law scaling and universal dynamics. Here, we exploit these regularities, using them to organize and unify social and physical properties of armed conflict in a scaling framework captured by our RBAC model.

Both qualitative understanding of conflict causes and observed regularities in the data motivate our starting assumption that multiple features of armed conflict abide by simple scaling laws \cite{richardsonVariationFrequency1948,clausetFrequencySevere2007,gillespieEstimatingNumber2017,johnsonSimpleMathematical2013,picoliUniversalBursty2015}. Although some of these features like the distribution of conflict sites might reflect a process external to conflict dynamics such as socioeconomic variability \cite{corralFragilityConflict2020}, it remains an open question of how such statistical patterns emerge in the first place. One set of hypotheses revolves around the idea that conflict is an example of self-organized criticality (SOC) \cite{jensenSelforganizedCriticality1998}. Roughly speaking, one might imagine that slow growth of social tension contrasted with relatively abrupt conflict resolution leads to scale-free features \cite{cedermanModelingSize2003}. This is a debated hypothesis, but we observe that SOC models such as forest fire models neither abide closely to our measured scaling laws nor account for the full set of conflict features \cite{cedermanModelingSize2003,robertsFractalitySelfOrganized1998,leeEmergentRegularities2019a}. At the least, SOC models must incorporate heterogeneity in space and time, which is, as we find, a defining feature of armed conflicts. Some physical analogs of these features like quenched disorder \cite{sethnaCracklingNoise2001,sethnaDeformationCrystals2017}, dissipation \cite{papanikolaouUniversalityPower2011}, or repetition on sites \cite{friedmanUniversalCritical2012,dickmanPathsSelforganized2000} have been considered in canonical models for criticality in nonequilibrium phenomena---though armed conflict suggests variations on these themes that may apply to social phenomena. More generally, the features we measure and the relations we establish between them in the RBAC model present a set of quantitative constraints that can be brought to bear on other models for armed conflict dynamics.

One constraint of particular note for conflict models results from our hypothesis that spatial scaling in armed conflict arises from the underlying geography on which it evolves \cite{richardsonStatisticsDeadly1960,zammit-mangionPointProcess2012}. As a way of capturing the fractal nature of conflict site density, we assume that conflict sites form a randomly branching tree. In this scenario, conflict features are determined by transportation networks, population density, and other social factors \cite{cohenDiffusionHomicide1999}. In intriguing alignment, some data suggest that the number of intersections of a road is characterized by a power law with exponent $2.2\leq u \leq 2.4$ \cite{kalapalaScaleInvariance2006}. Though conflict zones may be traversed in many ways, the overall statistics might be dominated by few major pathways such as the ring road in Afghanistan \cite{zammit-mangionPointProcess2012}. If so and if we think of intersections as meeting places where conflict actors converge, intersection density could account for why conflict extent is distributed with exponent $u=2.2$. Further support for the idea that transportation networks influence conflict comes from results showing fractal dimension of metropolitan road networks globally span the range $1.2\leq D\leq 1.7$ \cite{zhangFractalitySelfSimilarity2012}, findings that are in agreement with our exponent for armed conflict extent $\dn=1.6$. When a complete map of African transportation networks becomes available, it will be possible to further specify the mechanistic role of infrastructure on conflict dynamics.

Our approach reveals that conflict is not simply a geographic growth process but involves lattice-site dynamics resulting from its social nature. In particular, the density of reports and fatalities surpasses the two-dimensional physical landscape in which they are embedded, showing that the temporal dynamics at each lattice site are relevant. At each conflict site, reports and fatalities grow independently of geographic spread and are only rescaled in magnitude by final conflict duration. This suggests that conflict spreads locally in a common way---perhaps from shared social network structure across different parts of Africa or universal conflict spreading dynamics \cite{baudainsDynamicSpatial2016}. This would suggest that universality in conflict manifests in both local structure as well as in the statistics across many conflicts that span larger scales \cite{arcauteelsaCitiesRegions2016}. Overall, we find armed conflict dynamics are a consequence of underlying geography, asymmetry in between the core and periphery, and conflict virulence, aspects that are expressed through the scaling exponents.

Interestingly, our model reveals the presence of correlated fluctuations in conflict intensity, or conflict virulence, indicating spatiotemporal disorder separate from universal dynamics. Virulence specifically enhances fluctuations in social dimensions, reports and fatalities, in our model (though exponent differences suggest that some analog of virulence, e.g., population density, may matter for spatial extent, its effects are much weaker). Superlinear scaling of social phenomena with population number has been observed in the dynamics of cities and has been argued to promote innovation and growth \cite{bettencourtOriginsScaling2013}, but social scaling might likewise facilitate the spread of conflict, disinformation \cite{bakshyEveryoneInfluencer2011}, or disease \cite{maslovFlatteningCurve2020}. This aligns with the possibility that virulence reflects local social properties such as weak governance (e.g., comparing South Africa with Eastern Somalia \cite{besleyFragileStates2011}) or, similarly in primate societies, weak conflict management by leaders \cite{Flack:2006fk,FlaKraWaa05}. Alternatively, virulence could reflect a property of the instigating set of events as in primate society in which conflict duration grows with originating event severity \cite{leeCollectiveMemory2017}. Importantly, our finding of correlations in intensity over time suggests final conflict properties might be predicted at the onset. That conflict extinction is determined by the rate of events at the core is consistent with scaling in the data, suggesting that the origin of conflict outbreak is quantitatively, and perhaps plainly, linked to conflict duration.

Besides highlighting the importance of granular, high-resolution, and accurate social data to further the study of armed conflict \cite{cedermanPredictingArmed2017}, our work demonstrates the power of a thermodynamical approach to revealing and accounting for regularities in a complex and noisy social system \cite{johnsonSimpleMathematical2013}. If, as our minimal model suggests, geographic and social characteristics drive the evolution of conflict, then universality and scaling we observe may arise from the intersection of human social dynamics, regional properties like geography, and social structure.

\begin{acknowledgments}
We thank Guru Khalsa, Jaron Kent-Dobias, Van Savage, and Sid Redner for insightful discussion. EDL acknowledges funding from the Omega Miller Program, SFI Science, and Cornell University Graduate School. We acknowledge NSF no. 0904863 (JCF \& DCK), a St. Andrews Foundation grant of no. 13337 (EDL, JCF \& DCK), a John Templeton Foundation grant of no. 60501 (JCF \& DCK), the Proteus Foundation (JCF), and the Bengier Foundation (JCF).
\end{acknowledgments}
EDL, BCD, JCF, DCK contributed to ideation; EDL, BCD, and CRM constructed the model and performed the analysis; EDL and BCD took the lead in drafting the manuscript and all authors contributed to editing.

\clearpage
\appendix
\setcounter{figure}{0}
\setcounter{table}{0}
\renewcommand\thefigure{S.\arabic{figure}}
\renewcommand\thetable{S.\arabic{table}}

\section{Measuring conflict properties $\gamma_r$, $\gamma_f$, $V_r$, $V_f$}\label{sec:fitting}
Here, we describe how we measure the conflict site growth exponents $\gamma_r$ and $\gamma_f$ and the virulence $V_r$ and $V_f$.

\begin{figure}[tb]\centering
	\includegraphics[width=.85\linewidth]{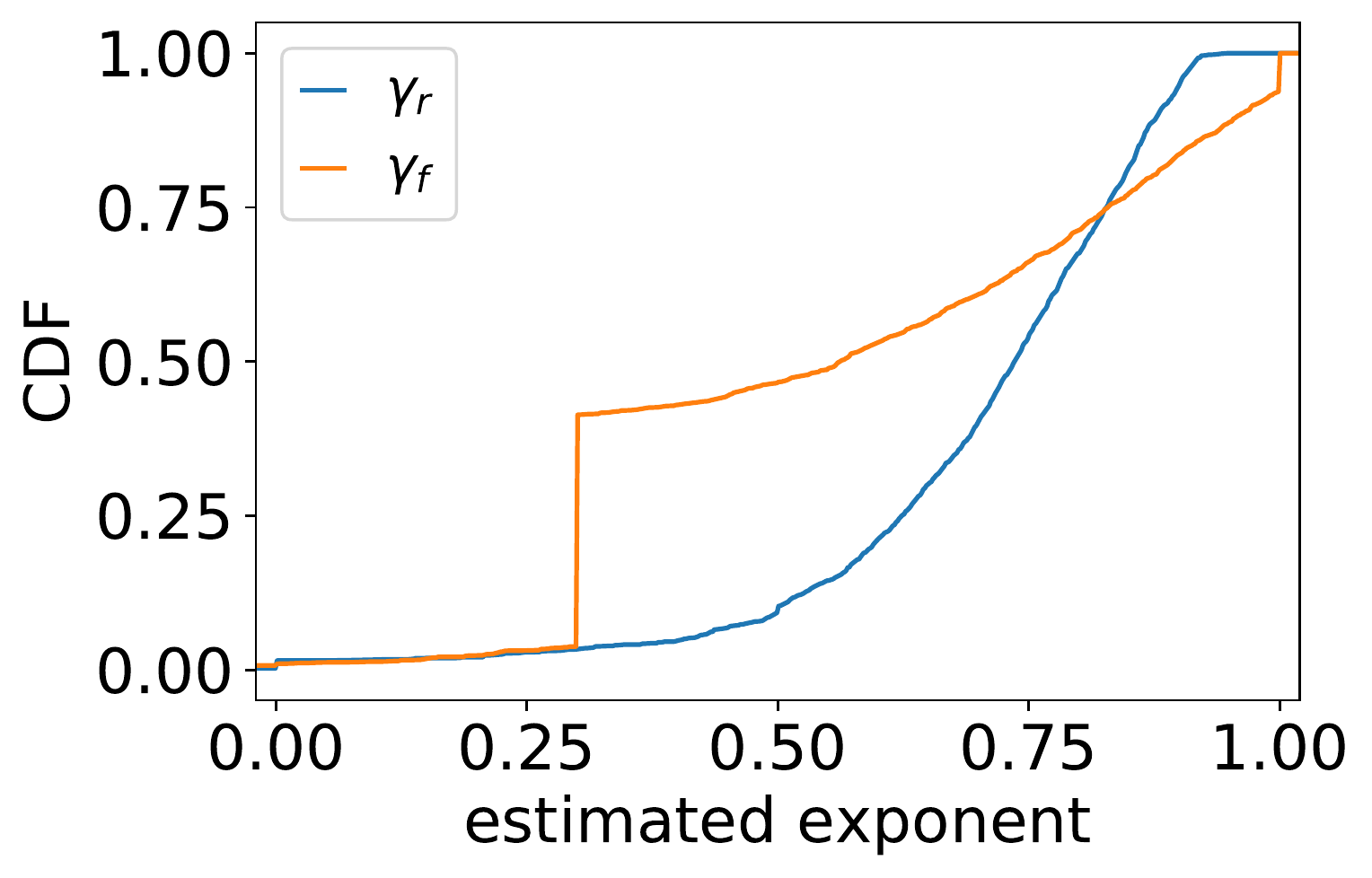}
	\caption{Cumulative distribution function (CDF) of exponents $\gamma_r$ and $\gamma_f$ estimated from regression to conflict site growth curves. Given this wide distribution, we take our best estimate of the exponent to be the median with confidence intervals given by the 5th and 95th percentiles as given in Table~\ref{tab:exponents}.}\label{gr:gamma estimate}
\end{figure}

We measure $\gamma_r$ by using the functional forms for site growth as in Eq~\ref{eq:log rxi}. To estimate the fitting parameters, we parameterize the logarithm of the scaling form to minimize the sum of two terms: one to fit the beginning of conflict avalanches and the other to fit the end. With reports as an example,
\begin{align}
\begin{aligned}
	&\argmin_{A,\gamma_r} \left\{\log[r_{x_i}(T)] - A +\right.\\
		&\quad\qquad\left.(1-\gamma_r)\log[T-t_0(x_i)+1]\right\}^2 + \\
		&\quad\qquad\{\log [r_{x_i}(0)] - A\}^2.
\end{aligned}
\end{align}
We constrain the sum $1-\gamma_r\geq0$. Then, we follow an analogous procedure for $\gamma_f$. The resulting distributions are shown in Figure~\ref{gr:gamma estimate}. Given the long tail we find, we use the medians as estimates of the exponents instead of the means.

Then, we take our best estimates of $\gamma_r$ and $\theta_r$, as described in the main text, to calculate the virulence per site at the end of the conflict avalanche, $t=T$. The averages of these measurements over all sites within a conflict avalanche returns the average $V_r$, which we show in Figure~\ref{gr:virulence}.

\section{Measuring $\theta_r$ and $\theta_f$}
\begin{figure}[tb]\centering
	\includegraphics[width=.85\linewidth]{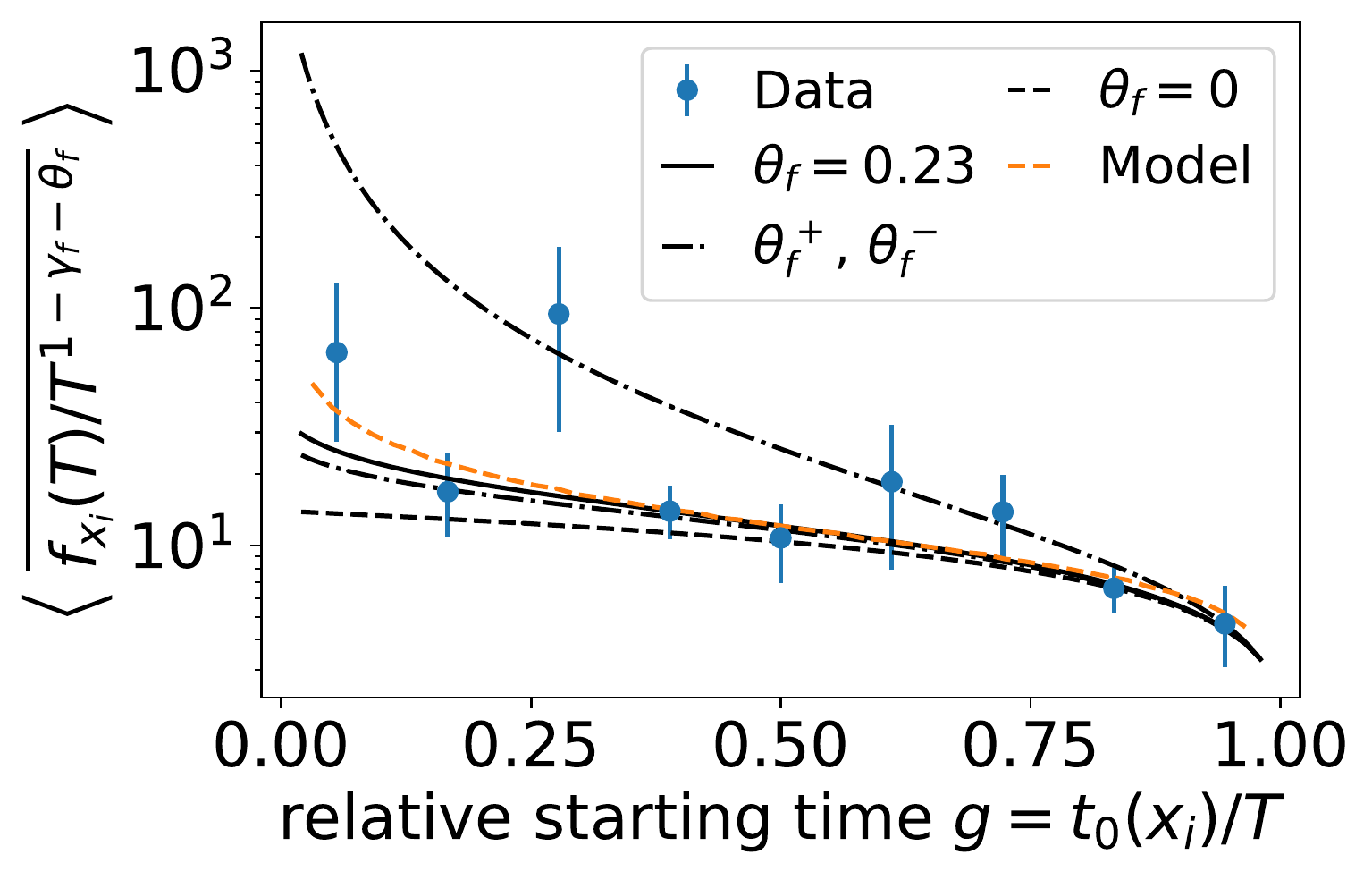}
	\caption{Scaling form predicted in Eq~\ref{eq:rxi scaling} given $\gamma_f=0.56$. We show bounds on $\theta_f$ corresponding to 90\% bootstrapped confidence intervals as $\theta_f^-$ and $\theta_f^+$. We compare with simulation (orange).}\label{gr:theta f}
\end{figure}

To measure the peripheral suppression exponents $\theta_r$ and $\theta_f$, we use the average profile defined in Eq~\ref{eq:rxi scaling}. We parameterize the fit to include a coefficient determining units $e^A$ and a small ``average'' correction $e^B$. The objective function for reports is the minimization problem
\begin{widetext}
\begin{align}
	\argmin_{\theta_r,A,B} \sum_{g} \sqrt{\left[\br{r_{x_i}(T)/T^{\theta_r+\gamma_r-1}} - e^A (1-g+e^B)^{1-\gamma_r}(g+e^B)^{-\theta_r}\right]^2/\sigma_{g}^2} + e^B,\label{eq:theta objective}
\end{align}
\end{widetext}
where the averaged profile for reports depends on $g$ implicitly through the relative time at which site $x_i$ started in the pertinent conflict avalanche. The form for $B$ ensures that it remain positive (or zero) and the second, cost term ensures that it remain small as is assumed in the derivation of the scaling form. The weighting terms $\sigma_g$ are the standard deviation of our measurements used to obtain the averaged profile $\br{r_{x_i}(T)/T^{\theta_r+\gamma_r-1}}$ such that the fit is more tightly constrained by the more precisely estimated points. Finally, we discretize the relative time $g\in[0,1]$ to intervals spaced out by $1/9$ as shown in Figures~\ref{gr:theta r} and \ref{gr:theta f}. We solve Eq~\ref{eq:theta objective} using standard optimization techniques \cite{2020SciPy-NMeth}. This procedure yields our initial estimates for the peripheral suppression exponents for the data.

For estimating the same exponents $\theta_r$ and $\theta_f$ from the RBAC simulation, however, there are two additional considerations that we take into account to solve the objective function defined in Eq~\ref{eq:theta objective}. First, we are able to obtain long conflict avalanches and the singularity at $t_0=0$ becomes important to consider. Indeed, if we fit the profile with the first point at relative time $g=0.056$, the emerging singularity at $g=0$ can substantially distort the measured value of $\theta_r$. On some test examples, we find that the point at $g=0.056$ jumps anomalously and forces the fit to match the remaining points poorly, an indication that our coarse-graining of $g$ into intervals of $1/9$ provides insufficient resolution to estimate $\theta_r$ accurately when avalanches are much longer than typical ones in the data. However, it is the case that far from $g=0$, the singularity has much smaller effect and by simply excluding the point at $g=0.056$, we recover accurately $\theta_r=0.5$, the value cited in the main text. Though in principle similar bias is also an issue for $g=1$, it does not skew our estimate of the exponents strongly and so we include it to replicate the procedure we use for the data as closely as possible. The second modification we make to the fitting procedure comes from the fact that $\sigma_g$ is no longer dominated by sampling noise and reflects the fact that fluctuations become larger near the singularity at $g=0$. Since fluctuations in the model are a function $\theta_r$, the objective behaves deterministically with $\theta_r$, $\theta_r$ is driven to large values, and the objective minimized by simply compressing the scaling function to vanishingly small values. For fitting the model, we replace $\sigma_g$ with $e^A$ such that the objective is rescaled by the typical value across the profile. We find that this allows us to get much more reasonable estimates for $\theta_r$ and $\theta_f$ while accounting for the typical scale of the average profile. Importantly, we find that these procedures lead to close fits of the averaged profile over the values of $g$ that we consider. Putting these pieces together, we find close agreement between the exponents estimated from the model and data, providing a way of confirming the validity of our fitting procedures using the model.

\begin{table*}[p]\centering
\caption{Reference for scaling variables and exponents. Questions marks stand in for an arbitrary scaling variable.}\vspace{10pt}
\begin{tabular}{c|l}
	Variable & Definition \\
\hline
	$\alpha$	& duration distribution exponent $P(T)\sim T^{-\alpha}$\\
	$\beta_?$	& distribution scaling exponent for virulence $P(V_?) \sim V_?^{-\beta_?}$\\
	$\gamma_?$	& exponent for conflict site endemicity\\
	$\delta_?$	& fractal dimension from $?(t)/?(T) \propto (l/L)^{\delta_?}$\\
	$\delta_?/\zeta$	& dynamical exponent for $?(t)/?(T) \propto (t/T)^{\delta_?/\zeta}$\\
	$\Delta_?$ 	& part of virulence dynamical scaling exponent $V_? \sim T^{\gamma_?+\Delta_?}$\\
	$\epsilon$	& correction to singularity in site growth\\
	$\zeta$		& dynamical exponent for diameter profile $l(t)/L \propto (t/T)^{1/\zeta}$\\
	$\theta_?$	& peripheral suppression exponent\\
	$\Lambda$	& maximum cutoff length for conflict region\\
	$\nu$		& diameter distribution exponent $P(L)\sim L^{-\nu}$\\	
	$\xi$		& correlation length\\
	$\tau$		& fatalities distribution exponent $P(F)\sim F^{-\tau}$\\
	$\tau'$		& reports distribution exponent $P(R)\sim R^{-\tau'}$\\
	$B$			& conflict tree branch extension ratio\\
	$C$			& reports threshold for conflict extinction\\
	$d_?/z$ 	& dynamical exponent for $?\propto T^{d_?/z}$\\
	$f$, $F$	& fatalities\\
	$l$, $L$    & diameter in km\\
	$n$, $N$    & number of conflict sites \\
	$Q$			& conflict tree branching number\\
	$r$, $R$    & number of reports\\
	$r_{x_i}$	& number of reports for site $x_i$\\
	$t_0(x_i)$	& time of first event at conflict site $x_i$\\
	$t$, $T$	& duration in days\\
	$u$			& extent distribution exponent $P(N)\sim N^{-u}$\\
	$v_?(x_i)$	& virulence for site $x_i$\\
	$V_?(x)$	& virulence averaged over sites $i$ in conflict avalanche $x$\\
	$\overline{V}_?$ & virulence averaged over different conflict avalanches\\
	$x$			& conflict avalanche with sites $\{x_i\}$\\
	$x_i$       & conflict site $i$ in conflict avalanche $x$\\
	$z$		    & dynamical exponent for length $L\propto T^{1/z}$\\
\end{tabular}\label{tab:variables}
\end{table*}

\section{RBAC simulation}\label{sec:simulation}
We start by growing a randomly branching tree of fractal dimension $\dn = 1.6$ (calculated from taking the ratio of the separately measured exponents $\dn/\zeta$ and $1/\zeta$) emanating from a single seed site. Here, we consider $Q=3$ and produce an initial set of three branches with an average extension factor $B=6.6$. At each branching point, each set of children branches have random length $B^k(1+\eta)$, where $\eta$ is a random number chosen uniformly in the interval $[-\sigma_\eta, \sigma_\eta]$, $\sigma_\eta<1$ such that branches vary in length about the mean with fluctuations that grow proportionally with the mean. Given the lengths, the angle at which the branches split are chosen such that no branches will intersect with any other branches for a tree of arbitrary size. Examples of such random trees are shown in Figure~\ref{gr:ntd examples}.

On every newly added site, report and fatality dynamics are instigated such that the total number of events grow according to Eq~\ref{eq:site scaling}. We set site dynamical exponents to their best fits: $\gamma_r=0.74$, $\theta_r=0.43$, $\gamma_f=0.56$, $\theta_f=0.23$, with $V_r$ sampled from power law distribution with exponent $\beta_r=3$ and lower bound of $V_{r,0}=1$ to avoid very small conflict avalanches dominated by finite-size effects. At each conflict site, we treat the total cumulative number of events to be a continuous function of the discrete number of time steps $t_0(x_i)$ as would be the case in the limit of large avalanches.\footnote{Discretization of the continuous measures of reports and fatalities introduces finite-size effects that become unimportant for large avalanches. Though we do not necessarily expect that the corrections introduced by discretization of our conflict avalanches align with those in the data, this issue represents a question of interest for future work that grapples with deviations from scaling.} This gives us the trajectories per site $r_{x_i}(t)$ and conflict avalanche evolution $r_x(t)$ as well as the corresponding trajectories for fatalities, $f_{x_i}(t)$ and $f_x(t)$.

Conflict avalanches are run til they reach the threshold rate of events determined by the scaling relation in Eq~\ref{eq:V_r for constant threshold}. To simulate this, we take the random sample for virulence $V_r$ as mentioned above. Given a fixed, universal conflict rate threshold (e.g., $C=2^{-7}$, or one event per 128 days), the simulation ends when the mean event rate at the core crosses the threshold
\begin{align}
	\left.\frac{\partial r_{x_{\rm i}}}{\partial t}\right|_{t_0=0} &= (1-\gamma_r) V_r(x) (t+1)^{-\gamma_r}.
\end{align}
Thus, conflict extinction is determined by the combination of our fixed threshold for conflict rate, conflict avalanche virulence, and the universal rate with which it decays. The results are shown in Figure~\ref{gr:rbac model}.

\clearpage
\bibliography{refs}

\begin{thebibliography}{10}

\bibitem{richardsonVariationFrequency1948}
Lewis~F. Richardson.
\newblock Variation of the {{Frequency}} of {{Fatal Quarrels With Magnitude}}.
\newblock {\em J. Am. Stat. Assoc.}, 43(244):523--546, 1948.

\bibitem{clausetPowerLawDistributions2009}
Aaron Clauset, Cosma~Rohilla Shalizi, and Mark E.~J. Newman.
\newblock Power-{{Law Distributions}} in {{Empirical Data}}.
\newblock {\em SIAM Rev.}, 51(4):661--703, November 2009.

\bibitem{clausetFrequencySevere2007}
Aaron Clauset, Maxwell Young, and Kristian~Skrede Gleditsch.
\newblock On the {{Frequency}} of {{Severe Terrorist Events}}.
\newblock {\em J. Confl. Resolut.}, 51(1):58--87, February 2007.

\bibitem{gillespieEstimatingNumber2017}
Colin~S. Gillespie.
\newblock Estimating the number of casualties in the {{American Indian}} war:
  {{A Bayesian}} analysis using the power law distribution.
\newblock {\em Ann. Appl. Stat.}, 11(4):2357--2374, December 2017.

\bibitem{johnsonSimpleMathematical2013}
Neil~F. Johnson, Pablo Medina, Guannan Zhao, Daniel~S. Messinger, John Horgan,
  Paul Gill, Juan~Camilo Bohorquez, Whitney Mattson, Devon Gangi, Hong Qi,
  Pedro Manrique, Nicolas Velasquez, Ana Morgenstern, Elvira Restrepo, Nicholas
  Johnson, Michael Spagat, and Roberto Zarama.
\newblock Simple mathematical law benchmarks human confrontations.
\newblock {\em Sci. Rep.}, 3(1):3463, December 2013.

\bibitem{picoliUniversalBursty2015}
S.~Picoli, M.~del {Castillo-Mussot}, H.~V. Ribeiro, E.~K. Lenzi, and R.~S.
  Mendes.
\newblock Universal bursty behaviour in human violent conflicts.
\newblock {\em Sci. Rep.}, 4(1):4773, May 2015.

\bibitem{raleighIntroducingACLED2010}
Clionadh Raleigh, Andrew Linke, H{\aa}vard Hegre, and Joakim Karlsen.
\newblock Introducing {{ACLED}}: {{An Armed Conflict Location}} and {{Event
  Dataset}}: {{Special Data Feature}}.
\newblock {\em J. Peace Res.}, 47(5):651--660, September 2010.

\bibitem{leeEmergentRegularities2019a}
Edward~D. Lee, Bryan~C. Daniels, Christopher~R. Myers, David~C. Krakauer, and
  Jessica~C. Flack.
\newblock Emergent regularities and scaling in armed conflict data.
\newblock {\em arXiv:1903.07762 [cond-mat, physics:nlin, physics:physics,
  q-bio]}, October 2019.

\bibitem{sethnaDeformationCrystals2017}
James~P. Sethna, Matthew~K. Bierbaum, Karin~A. Dahmen, Carl~P. Goodrich,
  Julia~R. Greer, Lorien~X. Hayden, Jaron~P. {Kent-Dobias}, Edward~D. Lee,
  Danilo~B. Liarte, Xiaoyue Ni, Katherine~N. Quinn, Archishman Raju, D.~Zeb
  Rocklin, Ashivni Shekhawat, and Stefano Zapperi.
\newblock Deformation of {{Crystals}}: {{Connections}} with {{Statistical
  Physics}}.
\newblock {\em Annu. Rev. Mater. Res.}, 47(1):217--246, 2017.

\bibitem{leeCollectiveMemory2017}
Edward~D. Lee, Bryan~C. Daniels, David~C. Krakauer, and Jessica~C. Flack.
\newblock Collective memory in primate conflict implied by temporal scaling
  collapse.
\newblock {\em J. Royal Soc. Interface}, 14(134):20170223, September 2017.

\bibitem{fortunatoScalingUniversality2007}
Santo Fortunato and Claudio Castellano.
\newblock Scaling and {{Universality}} in {{Proportional Elections}}.
\newblock {\em Phys. Rev. Lett.}, 99(13):138701, September 2007.

\bibitem{bettencourtOriginsScaling2013}
L.~M.~A. Bettencourt.
\newblock The {{Origins}} of {{Scaling}} in {{Cities}}.
\newblock {\em Science}, 340(6139):1438--1441, June 2013.

\bibitem{castellanoStatisticalPhysics2009}
Claudio Castellano, Santo Fortunato, and Vittorio Loreto.
\newblock Statistical physics of social dynamics.
\newblock {\em Rev. Mod. Phys.}, 81(2):591--646, May 2009.

\bibitem{newmanSpreadEpidemic2002}
M.~E.~J. Newman.
\newblock Spread of epidemic disease on networks.
\newblock {\em Phys. Rev. E}, 66(1):016128, July 2002.

\bibitem{poncealvarezWholeBrainNeuronal2018}
Adri{\'a}n {Ponce-Alvarez}, Adrien Jouary, Martin Privat, Gustavo Deco, and
  Germ{\'a}n Sumbre.
\newblock Whole-{{Brain Neuronal Activity Displays Crackling Noise Dynamics}}.
\newblock {\em Neuron}, 100(6):1446--1459.e6, December 2018.

\bibitem{friedmanUniversalCritical2012}
Nir Friedman, Shinya Ito, Braden A.~W. Brinkman, Masanori Shimono, R.~E.~Lee
  DeVille, Karin~A. Dahmen, John~M. Beggs, and Thomas~C. Butler.
\newblock Universal {{Critical Dynamics}} in {{High Resolution Neuronal
  Avalanche Data}}.
\newblock {\em Phys. Rev. Lett.}, 108(20):208102, May 2012.

\bibitem{fonteneleCriticalityCortical2019}
Antonio~J. Fontenele, Nivaldo A.~P. {de Vasconcelos}, Tha{\'i}s Feliciano,
  Leandro A.~A. Aguiar, Carina {Soares-Cunha}, B{\'a}rbara Coimbra, Leonardo
  Dalla~Porta, Sidarta Ribeiro, Ana~Jo{\~a}o Rodrigues, Nuno Sousa, Pedro~V.
  Carelli, and Mauro Copelli.
\newblock Criticality between {{Cortical States}}.
\newblock {\em Phys. Rev. Lett.}, 122(20):208101, May 2019.

\bibitem{cedermanModelingSize2003}
Lars-Erik Cederman.
\newblock Modeling the {{Size}} of {{Wars}}: {{From Billiard Balls}} to
  {{Sandpiles}}.
\newblock {\em APSR}, 97(01):135--150, February 2003.

\bibitem{osullivanDominoesDice1996}
Patrick O'Sullivan.
\newblock Dominoes or {{Dice}}: {{Geography}} and the {{Diffusion}} of
  {{Political Violence}}.
\newblock {\em JCS}, 16(2), 1996.

\bibitem{schutteDiffusionPatterns2011}
Sebastian Schutte and Nils~B. Weidmann.
\newblock Diffusion patterns of violence in civil wars.
\newblock {\em Political Geogr.}, 30(3):143--152, March 2011.

\bibitem{corralFragilityConflict2020}
Paul Corral, Alexander Irwin, Nandini Krishnan, Daniel~G. Mahler, and Tara
  Vishwanath.
\newblock {\em Fragility and {{Conflict}}: {{On}} the {{Front Lines}} of the
  {{Fight Against Poverty}}}.
\newblock {World Bank Group}, 2020.

\bibitem{kalapalaScaleInvariance2006}
Vamsi Kalapala, Vishal Sanwalani, Aaron Clauset, and Cristopher Moore.
\newblock Scale invariance in road networks.
\newblock {\em Phys. Rev. E}, 73(2):026130, February 2006.

\bibitem{marshallGlobalReport2008}
Monty~G. Marshall and Benjamin~R. Cole.
\newblock Global {{Report}} on {{Conflict}}, {{Governance}} and {{State
  Fragility}} 2008.
\newblock {\em Foreign Pol. Bull.}, 18(1):3--21, 2008.

\bibitem{hussainOpeningClosed2012}
Muzammil~M. Hussain and Philip~N. Howard.
\newblock Opening {{Closed Regimes}}.
\newblock In Eva Anduiza, Michael~James Jensen, and Laia Jorba, editors, {\em
  Digital {{Media}} and {{Political Engagement Worldwide}}}, pages 200--220.
  {Cambridge University Press}, {Cambridge}, 2012.

\bibitem{sealeStruggleSyria1987}
Patrick Seale.
\newblock {\em The {{Struggle}} for {{Syria}}: {{A}} Study of {{Post}}-{{War
  Arab Politics}}, 1945-1958}.
\newblock {Yale University Press}, 1987.

\bibitem{burioniFractalsAnomalous1994}
Raffaella Burioni and Davide Cassi.
\newblock Fractals without anomalous diffusion.
\newblock {\em Phys. Rev. E}, 49(3):R1785--R1787, March 1994.

\bibitem{oloughlinPeeringFog2010}
John O'Loughlin, Frank D.~W. Witmer, Andrew~M. Linke, and Nancy Thorwardson.
\newblock Peering into the {{Fog}} of {{War}}: {{The Geography}} of the
  {{WikiLeaks Afghanistan War Logs}}, 2004-2009.
\newblock {\em Eurasian Geogr. Econ.}, 51(4):472--495, July 2010.

\bibitem{lynchArabUprising2013}
Mark Lynch.
\newblock {\em The {{Arab Uprising}}: {{The Unfinished Revolutions}} of the
  {{New Middle East}}}.
\newblock {PublicAffairs}, 2013.

\bibitem{jensenSelforganizedCriticality1998}
Henrik~Jeldtoft Jensen.
\newblock {\em Self-Organized Criticality: {{Emergent}} Complex Behavior in
  Physical and Biological Systems}.
\newblock Cambridge Lecture Notes in Physics. {Cambridge University Press},
  {Cambridge}, 1998.

\bibitem{robertsFractalitySelfOrganized1998}
D.~C. Roberts and D.~L. Turcotte.
\newblock Fractality and {{Self}}-{{Organized Criticality}} of {{Wars}}.
\newblock {\em Fractals}, 6(4):351--357, 1998.

\bibitem{sethnaCracklingNoise2001}
James~P. Sethna, Karin~A. Dahmen, and Christopher~R. Myers.
\newblock Crackling noise.
\newblock {\em Nature}, 410(6825):242, March 2001.

\bibitem{papanikolaouUniversalityPower2011}
Stefanos Papanikolaou, Felipe Bohn, Rubem~Luis Sommer, Gianfranco Durin,
  Stefano Zapperi, and James~P. Sethna.
\newblock Universality beyond power laws and the average avalanche shape.
\newblock {\em Nature Phys.}, 7(4):316--320, April 2011.

\bibitem{dickmanPathsSelforganized2000}
Ronald Dickman, Miguel~A. Mu{\~n}oz, Alessandro Vespignani, and Stefano
  Zapperi.
\newblock Paths to self-organized criticality.
\newblock {\em Braz. J. Phys.}, 30(1):27--41, March 2000.

\bibitem{richardsonStatisticsDeadly1960}
L.F. Richardson.
\newblock {\em Statistics of Deadly Quarrels}.
\newblock {Boxwood Press}, 1960.

\bibitem{zammit-mangionPointProcess2012}
A.~{Zammit-Mangion}, M.~Dewar, V.~Kadirkamanathan, and G.~Sanguinetti.
\newblock Point process modelling of the {{Afghan War Diary}}.
\newblock {\em Proc. Natl. Acad. Sci. U.S.A.}, 109(31):12414--12419, July 2012.

\bibitem{cohenDiffusionHomicide1999}
Jacqueline Cohen and George Tita.
\newblock Diffusion in {{Homicide}}: {{Exploring}} a {{General Method}} for
  {{Detecting Spatial Diffusion Processes}}.
\newblock {\em J. Quant. Criminol.}, 15(4):43, 1999.

\bibitem{zhangFractalitySelfSimilarity2012}
Hong Zhang and Zhilin Li.
\newblock Fractality and {{Self}}-{{Similarity}} in the {{Structure}} of {{Road
  Networks}}.
\newblock {\em Ann. Am. Assoc. Geogr.}, 102(2):350--365, March 2012.

\bibitem{baudainsDynamicSpatial2016}
P.~Baudains, H.M. Fry, T.P. Davies, A.G. Wilson, and S.R. Bishop.
\newblock A dynamic spatial model of conflict escalation.
\newblock {\em Eur. J. Appl. Math}, 27(3):530--553, June 2016.

\bibitem{arcauteelsaCitiesRegions2016}
{Arcaute Elsa}, {Molinero Carlos}, {Hatna Erez}, {Murcio Roberto}, {Vargas-Ruiz
  Camilo}, {Masucci A. Paolo}, and {Batty Michael}.
\newblock Cities and regions in {{Britain}} through hierarchical percolation.
\newblock {\em Royal Soc. Open Sci.}, 3(4):150691, 2016.

\bibitem{bakshyEveryoneInfluencer2011}
Eytan Bakshy, Jake~M. Hofman, Winter~A. Mason, and Duncan~J. Watts.
\newblock Everyone's an influencer: {{Quantifying}} influence on {{Twitter}}.
\newblock In {\em Proceedings of the Fourth {{ACM}} International Conference on
  {{Web}} Search and Data Mining}, page~65, {Hong Kong, China}, 2011. {ACM
  Press}.

\bibitem{maslovFlatteningCurve2020}
Sergei Maslov and Nigel Goldenfeld.
\newblock Flattening the curve fails unless done very early: Results from a
  simulation of {{ICU}} capacity in {{Chicago}}.
\newblock March 2020.

\bibitem{besleyFragileStates2011}
Timothy Besley and Torsten Persson.
\newblock Fragile {{States}} and {{Development Policy}}.
\newblock {\em J. Eur. Econ. Assoc.}, 9(3):371--398, June 2011.

\bibitem{Flack:2006fk}
Jessica~C. Flack, Michelle Girvan, Frans B.~M. {de Waal}, and David~C.
  Krakauer.
\newblock Policing stabilizes construction of social niches in primates.
\newblock {\em Nature}, 439(7075):426--429, January 2006.

\bibitem{FlaKraWaa05}
Jessica~C Flack, David~C Krakauer, and Frans B~M {de Waal}.
\newblock Robustness mechanisms in primate societies: A perturbation study.
\newblock {\em Proc. R. Soc. B}, 272(1568):1091--9, June 2005.

\bibitem{cedermanPredictingArmed2017}
Lars-Erik Cederman and Nils~B. Weidmann.
\newblock Predicting armed conflict: {{Time}} to adjust our expectations?
\newblock {\em Science}, 355(6324):474--476, February 2017.

\bibitem{2020SciPy-NMeth}
Pauli Virtanen, Ralf Gommers, Travis~E. Oliphant, Matt Haberland, Tyler Reddy,
  David Cournapeau, Evgeni Burovski, Pearu Peterson, Warren Weckesser, Jonathan
  Bright, St{\'e}fan~J. {van der Walt}, Matthew Brett, Joshua Wilson,
  K.~Jarrod~Millman, Nikolay Mayorov, Andrew R.~J. Nelson, Eric Jones, Robert
  Kern, Eric Larson, CJ~Carey, {\.I}lhan Polat, Yu~Feng, Eric~W. Moore, Jake
  {Vand erPlas}, Denis Laxalde, Josef Perktold, Robert Cimrman, Ian Henriksen,
  E.~A. Quintero, Charles~R Harris, Anne~M. Archibald, Ant{\^o}nio~H. Ribeiro,
  Fabian Pedregosa, Paul {van Mulbregt}, and SciPy 1.~0 Contributors.
\newblock {{SciPy}} 1.0: {{Fundamental}} algorithms for scientific computing in
  python.
\newblock {\em Nature Methods}, 17:261--272, 2020.

\end{thebibliography}

\end{document}